\documentclass[10pt,journal,compsoc]{IEEEtran}

\ifCLASSOPTIONcompsoc
  \usepackage[nocompress]{cite}
\else
  \usepackage{cite}
\fi
\ifCLASSINFOpdf
\else
\fi

\usepackage{boldline}
\usepackage[x11names, table]{xcolor}
\usepackage{colortbl}
\usepackage{bigstrut}
\usepackage[ruled]{algorithm2e}
\usepackage{amsmath,amssymb,amsfonts}
\usepackage{amsmath}
\usepackage{array}
\usepackage{algorithmic}
\usepackage{balance}
\usepackage{blindtext}
\usepackage{booktabs}
\usepackage{caption}
\usepackage{hyperref}
\usepackage[noabbrev]{cleveref}
\usepackage{cite}
\usepackage{color}
\usepackage{comment}
\usepackage{enumitem}
\usepackage{eqparbox}
\usepackage{fancybox}
\usepackage{fancyvrb}
\usepackage{floatrow}
\usepackage{framed}
\usepackage{graphicx}
\usepackage{ifthen}
\usepackage{listings}
\usepackage{mathrsfs}
\usepackage{mdwmath}
\usepackage{mdwtab}
\usepackage{multirow}
\usepackage{pifont}
\usepackage{stfloats}
\usepackage{subfigure}
\usepackage{textcomp}
\usepackage{times}
\usepackage{url}
\usepackage{xcolor}
\usepackage{xspace}
\usepackage[normalem]{ulem}
\usepackage[framemethod=TikZ]{mdframed}

\newcommand{\subparagraph}{}
\usepackage{titlesec}

\usepackage[turnon]{notes}

 \usepackage{tcolorbox}
 \tcbuselibrary{listings,skins}
 
 \usepackage{floatrow}
\usepackage{mathtools}
\usepackage{blindtext}

\DeclareGraphicsExtensions{.pdf,.jpeg,.png}
\graphicspath{{figures/}}

\titlespacing\section{0pt}{8pt plus 4pt minus 2pt}{4pt plus 2pt minus 2pt}

\renewcommand{\baselinestretch}{0.99}
\floatstyle{plaintop}
\restylefloat{table}

\newcommand{\cmark}{\ding{51}}%
\newcommand{\xmark}{\ding{55}}%

\newcommand{\rom}[1]{\uppercase\expandafter{\romannumeral #1\relax}}

\newcommand{\etal}{\hbox{\emph{et al.}}\xspace}
\newcommand{\eg}{\hbox{\emph{e.g.,}}\xspace}
\newcommand{\ie}{\hbox{\emph{i.e.}}\xspace}

\newcommand{\wrt}{\hbox{\emph{w.r.t.}}\xspace}

\newcommand{\vs}{\hbox{\emph{v.s.}}\xspace}
\newcommand{\etc}{\hbox{\emph{etc}}\xspace}

\definecolor{gray50}{gray}{.5}
\definecolor{gray40}{gray}{.6}
\definecolor{gray30}{gray}{.7}
\definecolor{gray20}{gray}{.8}
\definecolor{gray10}{gray}{.9}
\definecolor{gray05}{gray}{.95}

\newlength\Linewidth
\def\findlength{\setlength\Linewidth\linewidth
\addtolength\Linewidth{-4\fboxrule}
\addtolength\Linewidth{-3\fboxsep}
}
\newenvironment{examplebox}{\par\begingroup
  \setlength{\fboxsep}{5pt}\findlength
  \setbox0=\vbox\bgroup\noindent
  \hsize=0.95\linewidth
  \begin{minipage}{0.95\linewidth}\normalsize}
    {\end{minipage}\egroup
    \textcolor{gray20}{\fboxsep1.5pt\fbox
     {\fboxsep5pt\colorbox{gray05}{\normalcolor\box0}}}
    \endgroup\par\noindent
    \normalcolor\ignorespacesafterend}

\newcounter{RQCounter}
\newcounter{RQACounter}

\newcommand{\RQ}[2]{%
\refstepcounter{RQACounter} \label{#1}
\noindent\textbf{RQ\arabic{RQACounter}.~#2
}
}

\newcommand{\RS}[2]{%
\begin{mdframed}
\textbf{Result {\ref{#1}}:~}{\emph {#2}}%
\end{mdframed}
}

\definecolor{javared}{rgb}{0.6,0,0} 
\definecolor{javagreen}{rgb}{0.25,0.5,0.35} 
\definecolor{javapurple}{rgb}{0.5,0,0.35} 
\definecolor{javadocblue}{rgb}{0.25,0.35,0.75} 

\lstdefinestyle{customjava}{
  belowcaptionskip=\baselineskip,
  breaklines=true,
  language=java,
  showstringspaces=false,
  basicstyle=\small\ttfamily,
  keywordstyle=\bfseries\color{javapurple},
  commentstyle=\itshape\blue,
  identifierstyle=\blue,
  belowskip=-2pt,
  aboveskip=-2pt
}

\lstdefinestyle{codit}{
  belowcaptionskip=\baselineskip,
  breaklines=true,
  language=java,
  showstringspaces=false,
  basicstyle=\scriptsize\ttfamily,
  keywordstyle=\bfseries\color{javapurple},
  commentstyle=\itshape\blue,
  identifierstyle=\blue,
}

\newtcblisting{mylisting}[2][]{
    arc=3mm,
    listing only, 
    listing style=codit,
    title=#2,
    #1
    }

\DeclareMathOperator*{\argmax}{arg\,max}

\newcommand\Red[1]{\textcolor[rgb]{1.00,0.00,0.00}{\textbf{#1}}}

\newcommand\Orange[1]{\textcolor[rgb]{1.00,0.55,0.30}{{#1*}}}

\newcommand\blue[1]{\textcolor[rgb]{0.00,0.00,1.00}{{#1}}}
\newcommand\blued[1]{\textcolor[rgb]{0.00,0.00,1.00}{{#1$\dag$}}}

\newcommand\Green[1]{\textcolor[rgb]{0.0,0.6,0}{\textbf{#1}}}
\newcommand\Greenul[1]{\textcolor[rgb]{0.0,0.6,0}{{#1}}}

\newcommand\merun[1]{\textcolor[rgb]{0.45,0.0,0.0}{#1}}

\newcommand\javaKeyWord[1]{\textcolor[rgb]{0.38,0.03,0.58}{\tt \textbf{#1}}}

\definecolor{gray05}{gray}{0.95}
\definecolor{gray08}{gray}{0.92}
\definecolor{gray10}{gray}{0.90}
\definecolor{gray12}{gray}{0.88}
\definecolor{gray15}{gray}{0.85}
\definecolor{gray18}{gray}{0.82}
\definecolor{gray20}{gray}{0.80}
\definecolor{gray25}{gray}{0.75}
\definecolor{gray30}{gray}{0.70}
\definecolor{gray35}{gray}{0.65}
\definecolor{gray40}{gray}{0.60}
\definecolor{gray45}{gray}{0.55}
\definecolor{gray50}{gray}{0.50}
\definecolor{gray55}{gray}{0.45}
\definecolor{gray60}{gray}{0.40}
\definecolor{gray65}{gray}{0.35}
\definecolor{gray70}{gray}{0.30}
\definecolor{gray75}{gray}{0.25}
\definecolor{gray80}{gray}{0.20}
\definecolor{gray85}{gray}{0.15}
\definecolor{gray90}{gray}{0.10}
\definecolor{gray95}{gray}{0.05}

\newcommand{\ir}{$\mathscr{B}_{ir}$}

\newcommand{\tool}{\textsc{Codit}\xspace}

\xspace%

\newcommand{\cold}{$c_{p}$\xspace}
\newcommand{\cnew}{$c_{n}$\xspace}

\newcommand{\mold}{$m_{p}$\xspace}
\newcommand{\mnew}{$m_{n}$\xspace}
\newcommand{\vect}[1]{\ensuremath{\boldsymbol{#1}}}

\newcommand{\told}{$t_p$\xspace}
\newcommand{\tnew}{$t_n$\xspace}

\newcommand{\defj}{Defects4J\xspace}

\newcommand{\sts}{Seq2Seq\xspace}

\newcommand{\seq}{\sts}
\newcommand{\frontier}{{\tt frontier\_node}\xspace}

\newcommand{\Comment}[1]{}

\newcommand{\rqa}{{\xspace How accurately can \tool suggest concrete edits?}\xspace}
\newcommand{\rqb}{{\xspace How do different design choices affect \tool's performance?}\xspace}
\newcommand{\rqc}{{\xspace How accurately \tool suggests bug-fix patches?}\xspace}

\newcommand{\treemodel}{$\mathscr{M}_{tree}$\xspace}
\newcommand{\tokenmodel}{$\mathscr{M}_{token}$\xspace}

\newcommand{\dataname}{\textit{Code-Change-Data}\xspace}
\newcommand{\icsedata}{\textit{Pull-Request-Data}\xspace}

\newcommand{\ktree}{$K_{tree}$\xspace}
\newcommand{\ktoken}{$K_{token}$\xspace}

\begin{document}

\title{CODIT: Code Editing with Tree-Based Neural Models}

\author{Saikat Chakraborty,
        Yangruibo Ding,
        Miltiadis Allamanis,
        and~Baishakhi Ray%
        
\IEEEcompsocitemizethanks{
\IEEEcompsocthanksitem Saikat Chakraborty, Yangruibo Ding, and Baishakhi Ray are affiliated with Department of Computer Science, Columbia University, New York, NY 10027.\protect\\
E-mail: \{saikatc@cs., yangruibo.ding@, rayb@cs.\}columbia.edu
\IEEEcompsocthanksitem Miltiadis Allamanis is with Microsoft Research, Cambridge, UK.\protect\\
E-mail: miallama@microsoft.com.
}
}

\markboth{IEEE TRANSACTIONS ON SOFTWARE ENGINEERING, VOL. TBD, 2019}%
{Chakraborty \etal: CODIT: Code Editing with Tree-Based Neural Models}

\IEEEtitleabstractindextext{%


\begin{abstract}
The way developers edit day-to-day code tends to be repetitive, often using existing code elements. Many researchers have tried to automate repetitive code changes by learning from specific change templates which are applied to limited scope. The advancement of deep neural networks and the availability of vast open-source evolutionary data opens up the possibility of automatically learning those templates from the wild.  
However, deep neural network based modeling for code changes and code in general introduces some specific problems that needs specific attention from research community.
For instance, compared to natural language, source code vocabulary can be significantly larger. Further, good changes in code do not break its syntactic structure. Thus, deploying state-of-the-art neural network models without adapting the methods to the source code domain yields sub-optimal results.

To this end, we propose a novel tree-based neural network system to model source code changes and learn code change patterns from the wild. Specifically, we propose a tree-based neural machine translation model to learn the probability distribution of changes in code. We realize our model with a change suggestion engine, \tool, and train the model with more than 24k real-world changes
and evaluate it on 5k patches. Our evaluation shows the effectiveness of \tool in learning and suggesting patches. \tool can also learn specific bug fix pattern from bug fixing patches and can fix 25 bugs out of 80 bugs in \defj. 
\end{abstract}

}

\normalsize{%
  \setlength\abovedisplayskip{2pt}
  \setlength\belowdisplayskip{2pt}
  \setlength\abovedisplayshortskip{2pt}
  \setlength\belowdisplayshortskip{2pt}
}

\maketitle

\IEEEdisplaynontitleabstractindextext

\IEEEpeerreviewmaketitle



\IEEEraisesectionheading{\section{Introduction}\label{sec:intro}}

Developers edit source code to add new features, fix bugs, or maintain existing functionality (\eg API updates, refactoring, \etc.) all the time. Recent research has shown that these edits are often repetitive~\cite{nguyen2010recurring,nguyen2013study,ray2014uniqueness}. Moreover, the code components (\eg token, sub-trees, \etc.) used to build the edits are often taken from the existing codebase~\cite{martinez2014fix, barr2014plastic}. However, manually applying such repetitive edits can be tedious and error-prone~\cite{ray2013detecting}. Thus, it is important to automate code changes, as much as possible, to reduce the developers' burden. 

{There is a significant amount of industrial and academic work on automating code changes. For example, modern IDEs support specific types of automatic changes (\eg refactoring, adding boiler-plate code~\cite{vstudio,eclipse}, \etc). Many research tools aim to automate some types of edits, \eg API related changes~\cite{nguyen2010graph,tansey2008annotation,raychev2014code,andersen2012semantic,padioleau2008documenting,nguyen2016api}, refactoring~\cite{foster2012witchdoctor,raychev2013refactoring,ge2012reconciling,meng2015does}, frequently undergone code changes related to Pull Requests~\cite{tufano2019learning}, \etc.  Researchers have also proposed to automate generic changes by learning either from example edits~\cite{meng2011systematic,rolim2017learning} or from similar patches applied previously to source code~\cite{meng2013lase,nguyen2013study,ray2014uniqueness, tufano2019learning}. Automated {\em Code Change}\footnote{We use the term {\em Code Change} and {\em Code Edit} interchangeably} task is defined as modification of existing code (\ie, adding, deleting, or replacing code elements) through applying such frequent change patterns~\cite{ying2004predicting, zimmermann2005mining, kim2007automatic, tufano2019learning}}

While the above lines of work are promising and have shown initial success, they either rely on predefined change templates or require domain-specific knowledge: both are hard to generalize to the larger context. However, all of them leverage, in someway, common edit patterns. Given that a large amount of code and its change history is 
available thanks to software forges like GitHub, Bitbucket, \etc., 
a natural question arises: {\em Can we learn to predict general code changes by learning them in the wild?}  

Recently there has been a lot of interest in using Machine Learning (ML) techniques to model and predict source code from real world~\cite{allamanis2018survey}. However,  modeling changes is different from modeling generic code generation, since modeling changes is conditioned on the previous version of the code. In this work, we investigate whether ML models can capture the repetitiveness and statistical characteristics of code edits that developers apply daily. Such models should be able to automate code edits including feature changes, bug fixes, and other software evolution-related tasks.

A {\em code edit} can be represented as a tuple of two code versions:$< ${\em prev},{\em target}$ >$. 
To model the edits, one needs to learn the conditional probability distribution of the {\em target} code version given its {\em prev} code version. A good probabilistic model will assign higher probabilities to plausible target versions and lower probabilities to less plausible ones. 
Encoder-decoder Neural Machine Translation models (NMT) are a promising approach to realize such code edit models, where the previous code version is encoded into a latent vector representation. Then, the target version is synthesized (decoded) from the encoded representation. Indeed some previous works~\cite{tufano2019learning, tufano2018nmt_bug_fix, chen2018sequencer} use \seq NMT models to capture changes in token sequences. However, code edits also contain structural  changes, which need a syntax-aware model. 

To this end, we design a two step encoder-decoder model that models the probability distribution of changes.
In the first step, it learns to suggest structural changes in code using a tree-to-tree model, suggesting structural changes in the form of Abstract Syntax Tree (AST) modifications. 
Tree-based models, unlike their token-based counterparts, can capture the rich structure of code and always produce syntactically correct patches.
In the second step, the model concretizes the previously generated code fragment by predicting the tokens conditioned on the AST that was generated in the first step: given the type of each leaf node in the syntax tree, our model suggests concrete tokens of the correct type while respecting scope information. We combine these two models to realize \tool, a code change suggestion engine, which accepts a code fragment and generates potential edits of that snippet. 

{In this work, we particularly focus on smaller changes as our previous experience~\cite{ray2014uniqueness} shows that such changes mostly go through similar edits. In fact, all the previous NMT-based code transformation works~\cite{tufano2019learning, tufano2018nmt_bug_fix, chen2018sequencer} also aim to automate such changes. A recent study by Karampatsis \etal~\cite{RMKarampatsis2019SStuB} showed that small changes are  frequent\textemdash our analysis of the top 92 projects of their dataset found that, on average, in each project, one line changes take place around 26.71\% of the total commits and account for up to 70\% for the bug fix changes. Our focus in this work is primarily to automatically change small code fragments (often bounded by small AST sizes and/or few lines long) to reflect such repetitive patterns. Note that, in theory, our approach can be applied to any small fragment of code in the project repository with any programming language. However, for prototyping, we designed \tool to learn changes that belong to methods of popular java project in Github.}
In this work, we collect a new dataset~\textemdash~\dataname, consisting of 32,473 patches 
from 48 open-source GitHub projects collected from Travis Torrent~\cite{msr17challenge}. Our experiments show \tool achieves 15.94\% patch suggestion accuracy in the top 5 suggestions; this result outperforms a Copy-\sts baseline model by 63.34\% and a Tree2Seq based model by 44.37\%. 

We also evaluate \tool on \icsedata proposed by Tufano~\etal~\cite{tufano2019learning}. Our evaluation shows that \tool suggests 28.87\% of correct patches in the top 5 outperforming Copy-\sts-based model by 9.26\% and Tree2Seq based model by 22.92\%. 
Further evaluation on \tool's ability to suggest bug-fixing patches in \defj shows that \tool suggests 15 complete fixes and 10 partial fixes out of 80 bugs in \defj. 
%
%
%
In summary, our key contributions are: 
\begin{itemize}[leftmargin=*,label=-]
    \item 
We propose a novel tree-based code editing machine learning model that leverages the rich syntactic structure of code and generates syntactically correct patches. To our knowledge, we are the first to model code changes with tree-based machine translation.
\item 
We collect a large dataset of 32k real code changes. 
{Processed version of the dataset is available at \url{https://drive.google.com/file/d/1wSl_SN17tbATqlhNMO0O7sEkH9gqJ9Vr}}. 
\item 
We implement our approach, \tool, and evaluate the viability of using \tool for 
changes in a code, and patching bug fixes. Our code is available at 
{\url{https://git.io/JJGwU}}. 
\end{itemize}
%

\section{Background}
\label{sec:background}

\noindent
\textbf{Modeling Code Changes.}
Generating source code using machine learning models has been explored in the past~\cite{hindle2012naturalness,Tu:2014:FSE,hellendoorn2017deep,parvez2018building}. These methods model a probability distribution $p(c|\kappa)$ where $c$ is the generated code and $\kappa$ is any contextual information upon which the generated code is conditioned. In this work, we generate {\em code edits}. Thus, we are interested in models that predict code given its previous version. We achieve this using NMT-style models, which are a special case of  $p(c|\kappa)$, where $c$ is the new and $\kappa$ is the previous version of the code.  
NMT allows us to represent code edits with a single end-to-end model, taking into consideration the original version of a code and defining a conditional probability distribution of the target version. Similar ideas have been explored in NLP for paraphrasing~\cite{mallinson2017paraphrasing}.

\noindent
\textbf{Grammar-based modeling.}
Context Free Grammars (CFG) have been used to describe the syntax of programming languages~\cite{knuth1968semantics} and natural language~\cite{chomsky1956three, hopcroft2008introduction}. A CFG is a tuple $G = (N, \Sigma, P, S)$ where $N$ is a set of non-terminals, $\Sigma$ is a set of terminals, $P$ is a set of production rules in the form of $\alpha \rightarrow \beta$ and $a\in N$, $b \in (N \cup \Sigma)^*$, and $S$ is the start symbol. A sentence (\ie sequence of tokens) that belongs to the language defined by $G$ can be parsed by applying the appropriate derivation rules from the start symbol $S$. A common technique for generation of utterances is to expand the left-most, bottom-most non-terminal until all non-terminals have been expanded. Probabilistic context-free grammar (PCFG) is an extension of CFG, where each production rule in associated with a probability, \ie is defined as $(N, \Sigma, P, \Pi, S)$ where $\Pi$ defines a probability distribution for each production rule in $P$ conditioned on $\alpha$. 

\noindent
\textbf{Neural Machine Translation Models.}
NMT models are usually a cascade of an {\em encoder} and a {\em decoder}. The most common model is sequence-to-sequence (seq2seq)~\cite{bahdanau2014neural}, where the input is treated as a sequence of tokens and is encoded by a sequential encoder (\eg biLSTM). The output of the encoder is stored in an intermediate representation.
Next, the decoder using the encoder output and another sequence model, \eg an LSTM,
generates the output token sequence. In NMT, several improvements have been made over the base seq2seq model, such as {\em attention}~\cite{bahdanau2014neural} and {\em copying}~\cite{allamanis2016convolutional}. Attention mechanisms allow decoders to automatically search information within the input sequence when predicting each target element. Copying, a form of an attention mechanism, allows a model to directly copy elements of the input into the target. We employ both attention and copying mechanisms in this work. 



\section{Motivating Example}
\label{sec:motiv}

\begin{figure}[htpb]
 \subfigure[\textbf{Example of correctly suggested change by \tool along with the source and target parse trees. The deleted and added nodes are marked in~\Red{red} and~\Green{green} respectively.}]{
    \includegraphics[width=0.9\textwidth]{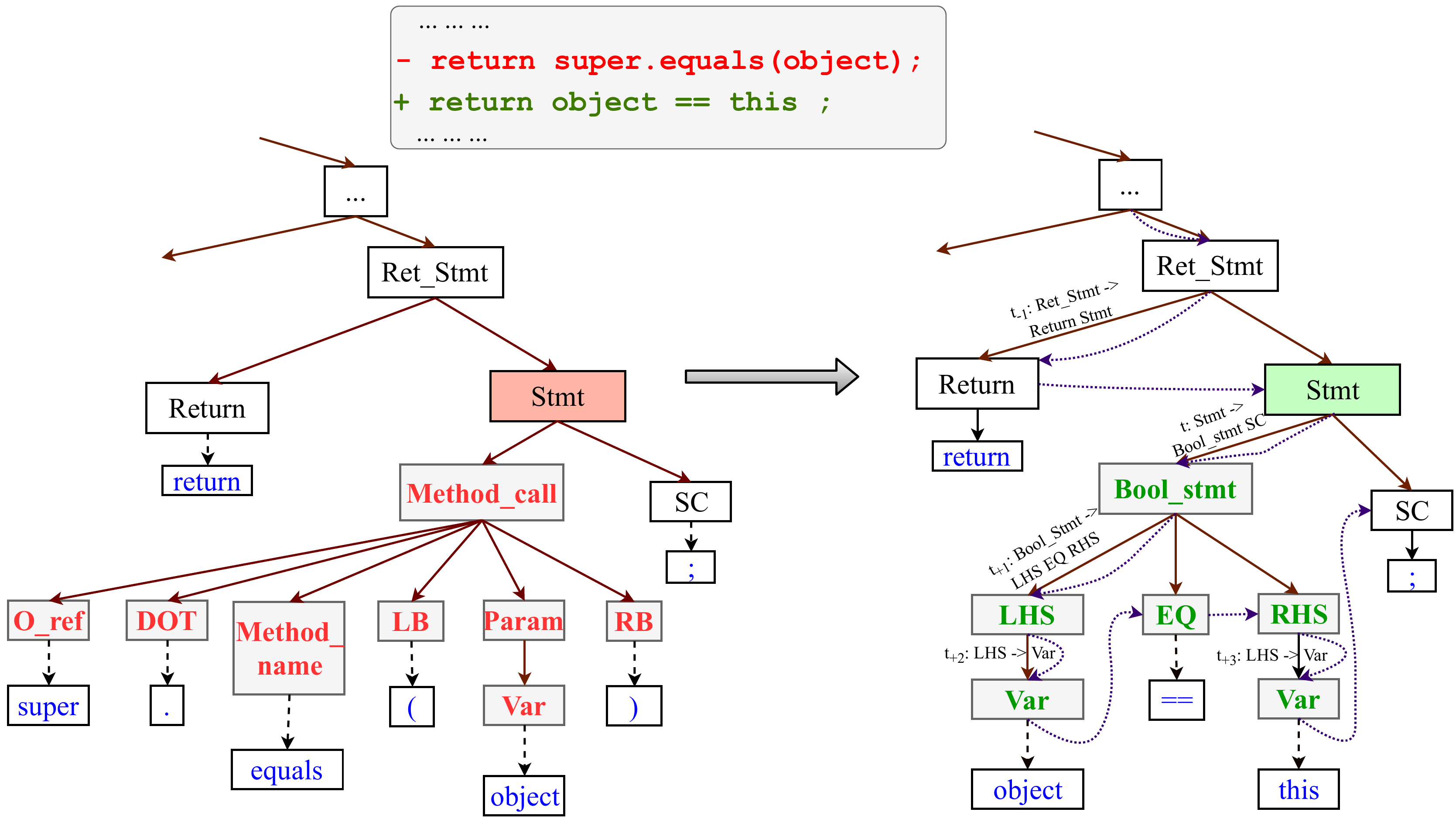}
    \label{fig:motiv_pt}
    }
\quad
\quad
\subfigure[\textbf{Sequence of grammar rules extracted from the parse trees.}]{
    \includegraphics[width=0.9\textwidth]{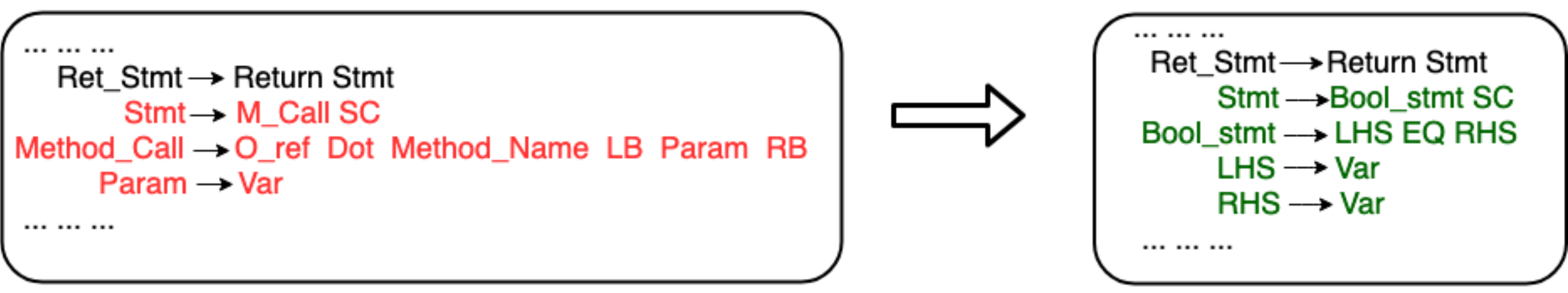}
    \label{fig:motiv_rules}
}
\quad
\quad
\medskip
\subfigure[\textbf{Token generation. Token probabilities are conditioned based on terminal types generated by tree translator (see \cref{fig:motiv_pt}) }]{
    \includegraphics[width=0.9\textwidth]{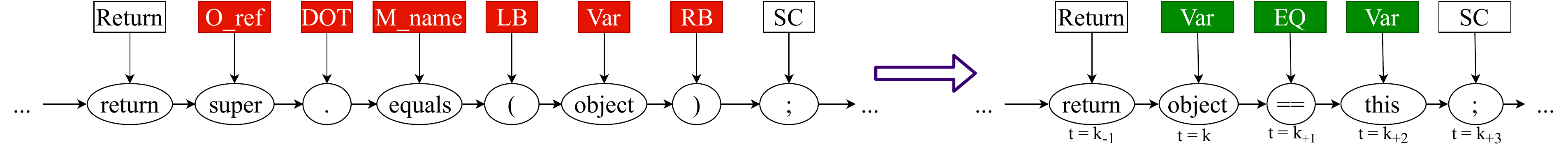}
    \label{fig:motiv_token}
}
\setlength{\belowcaptionskip}{-2cm} 
\caption{\textbf{\small Motivating Example}}
\label{fig:motiv}
\vspace{-10pt}
\end{figure}


~\Cref{fig:motiv} illustrates an example of our approach. 
Here, the original code fragment \Red{\tt return super.equals(object)} is edited to  \Green{\tt return object == this}.
\tool takes these two code fragments along with their context, for training. 
While suggesting changes, \ie, during test time, \tool takes as input the previous version of the code and generates its edited version.

\tool operates on the parse trees of the previous (\told) and new (\tnew) versions of the code, as shown in  
\Cref{fig:motiv_pt} (In the rest of the paper, a subscript or superscript with $p$ and $n$ correspond to \underline{p}revious and \underline{n}ew versions respectively). 
In~\Cref{fig:motiv}, changes are applied only to the subtree rooted at the $Method\_call$ node. The subtree is replaced by a new subtree (\tnew) with {\tt Bool\_stmt} as a root.
The deleted and added subtrees are highlighted in~\Red{red} and~\Green{green} respectively. 

While modeling the edit, \tool first predicts the structural changes in the parse tree. 
For example, in~\Cref{fig:motiv_pt} \tool first generates the changes corresponding to the subtrees with dark nodes and \merun{red} edges.
Next the structure is concretized by generating the token names (terminal nodes). 
This is realized by combining two models:
(i) a tree-based model predicting the structural change (see~\ref{sec:mtree}) followed by a 
(ii) a token generation model conditioned on the structure generated by the tree translation model (see~\ref{sec:mtoken}). 


\smallskip\noindent
\textit{Tree Translator}. The tree translator is responsible for generating structural changes to the tree structure.
A machine learning model is used to learn a (probabilistic) mapping between \told and \tnew.
First, a tree encoder, encodes \told computing a distributed vector representation for each of the production rules in \told yielding the distributed representation for the whole tree. Then, the tree decoder uses the encoded representations of \told to sequentially select rules from the language grammar to generate \tnew.
The tree generation starts with the {\tt root} node. Then, at each subsequent step, the bottom-most, left-most non-terminal node of the current tree is expanded. For instance, in~\Cref{fig:motiv_pt}, at time step {\tt t}, node {\tt Stmt} is expanded with rule {\tt \textbf{Stmt}} $\rightarrow$ \Green{\tt\textbf{Bool\_Stmt}} {\tt \textbf{SC}}.
When the tree generation process encounters a terminal node, it records the node type to be used by the token generation model and proceeds to the next non-terminal.
In this way, given the LHS rule sequences of \Cref{fig:motiv_rules} the RHS rule sequences is generated.


\smallskip\noindent
\textit{Token Generator}: The token generator predicts concrete tokens for the terminal node types generated in the previous step.
The token generator is a standard seq2seq model with attention and copying~\cite{bahdanau2014neural} but constrained on the token types generated by the tree translator. To achieve this, the token generator first encodes the token string representation and the node type sequence from \told.
The token decoder at each step probabilistically selects a token from the vocabulary or copies one from the input tokens in \told.
However, in contrast to traditional seq2seq where the generation of each token is only conditioned on the previously generated and source tokens, we additionally condition on the token type that has been predicted by the tree model and generate only tokens that are valid for that toke type. \Cref{fig:motiv_token} shows this step: given the token sequence of the original code {\tt < super . equals ( object ) >} and their corresponding token types (given in dark box), the new token sequence that is generated is {\tt < object == this >}.

\section{Tree-based Neural Translation Model}
\label{sec:model}

We decompose the task of predicting code changes in two stages: 
First, we learn and predict the structure (syntax tree) of the edited code. Then, given the predicted tree structure, we concretize the code. 
We factor the generation process as
\begin{align}
\vspace{-5mm}
    P(c_{n}|c_{p}) = P(c_n|t_n, c_p) P(t_n|t_p) P(t_p|c_p)
    \label{eqn:mainprob}
\end{align}
and our goal is to find $\hat{c}_{n}$ such that $\hat{c}_n\space =\space argmax_{c_n}P(c_n|c_p)$. 
Here, \cold is the previous version of the code and $t_p$ is its parse tree, whereas \cnew is the new version of the code and $t_n$ its parse tree.
Note that parsing a code fragment is unambiguous, \ie
$P(t_p|c_p) = 1$. Thus, our problem takes the form
\smallskip
\begin{equation}
\begin{split}
    \hat{c}_n = \argmax_{c_n, t_n} \underbrace{P(c_n | t_n, c_p)}_{\mathscr{M}_{token}} . \underbrace{P(t_n|t_p)}_{\mathscr{M}_{tree}}
\end{split}
    \label{eqn:final_prob_calc}
\end{equation}
Equation~\ref{eqn:final_prob_calc} has two parts. 
First, it estimates the changed syntax tree $P(t_n|t_p)$. We implement
this with a tree-based encoder-decoder model (\cref{sec:mtree}).
Next, given the predicted syntax tree $t_n$, we estimate the probability of the concrete edited code with $p(c_n|t_n, c_p)$ (\Cref{sec:mtoken}).

\subsection{Tree Translation Model (\treemodel)}
\label{sec:mtree}

The goal of \treemodel is to model the probability distribution of a new tree ($t_n$) given a previous version of the tree  ($t_p$). 
For any meaningful code the generated tree is syntactically correct. 
We represent the tree as a sequence of grammar rule generations following the CFG of the underlying programming language. The tree is generated by iteratively applying CFG expansions at the left-most bottom-most non-terminal node (\frontier) starting from the start symbol.  

For example, consider the tree fragments in~\Cref{fig:motiv_pt}. ~\Cref{fig:motiv_rules} shows the sequence of rules that generate those trees. For example, in the right tree of~\Cref{fig:motiv_pt}, the node {\tt Ret\_Stmt} is first  expanded by the rule: {\tt Ret\_Stmt$\rightarrow$Return Stmt}. 
Since, {\tt Return} is a terminal node, it is not expanded any further. Next, node {\tt Stmt} is expanded with rule: {\tt Stmt$\rightarrow$Bool\_Stmt SC}. The tree is further expanded with {
\tt Bool\_Stmt$\rightarrow$LHS EQ RHS}, 
{\tt LHS$\rightarrow$Var}, and {\tt RHS$\rightarrow$Var}. 
During the tree generation process, we apply these rules to yield the tree fragment of the next version.

In particular, the tree is generated by picking CFG rules at each non-terminal node. Thus, our model resembles a Probabilistic Context-Free Grammar (PCFG), but the probability of each rule depends on its surroundings.
The neural network models the probability distribution, $P(R^n_k | R_1^n, ... R_{k-1}^n, t_p)$: At time $k$ the probability of a rule depends on the input tree $t_p$ and the rules $R_1^n, ... R_{k-1}^n$ that have been applied so far.
Thus, the model for generating the syntax tree $t_n$ is given by
\begin{equation}
    P(t_n|t_p) = \prod\limits_{k=1}^{\tau} P(R_k^n|R_1^n, ... R_{k-1}^n, t_p) 
    \label{eqn:tree_prob}
\end{equation}

\medskip
\noindent
\textit{Encoder:}
The encoder encodes the sequence of rules that construct $t_p$.
For every rule $R_i^p$ in $t_p,$
we first transform it into a single learnable distributed vector representation $\vect{r}_{R_i^p}$. Then, the LSTM encoder summarizes the whole sequence up to position $i$ into a single vector $\vect{h}_i^p$. 
\begin{equation}
    \vect{h}_i^p = f_{LSTM}(\vect{h}_{i-1}^p, \vect{r}_{R_i^p})
    \label{eqn:encoder}
\end{equation}
This hidden vector contains information about the particular rule being applied and the previously applied rules. 
Once all the rules in $t_p$ are processed, we get a final hidden representation ($\vect{h}_\tau^p$).
The representations at each time step ($\vect{h}_1^p, \vect{h}_2^p, ..., \vect{h}_\tau^p)$ are used in the decoder to generate rule sequence for the next version of the tree. The parameters of the LSTM and the rules representations $\vect{r}_{R_i^p}$ are randomly initialized and learned jointly with all other model parameters.

\medskip
\noindent
\textit{Decoder:}
Our decoder has an LSTM with an attention mechanism as described by Bahdanau \etal~\cite{bahdanau2014neural}. The decoder LSTM is initialized with the final output from the encoder, \ie $ \vect{h}_0^n =\vect{h}_\tau^p$.
At a given decoding step $k$ the decoder LSTM changes its internal state in the following way, 
\smallskip
\begin{equation}
    \vect{h}_k^n = f_{LSTM}(\vect{h}_{k-1}^n, \vect{\psi}_k),
    \label{eqn:h_decoder}
\end{equation}
where $\vect{\psi}_k$ is computed by the attention-based weighted sum of the inputs $\vect{h}^p_j$ as \cite{bahdanau2014neural} in , \ie
\smallskip
\begin{equation}
   \vect{\psi}_k = \sum\limits_{j=1}^{\tau}
   softmax({\vect{h}_{k-1}^n}^T \vect{h}_j^p) \vect{h}_j^p
    \label{eqn:context}
\end{equation}
Then,
the probability over the rules at the $k$th step is:
\smallskip
\begin{multline}
    P(R_k^n| R_1^n, ... R_{k-1}^n, t_p) = softmax(W_{tree} \cdot \vect{h}_k^n + \mathbf{b}_{tree})
    \label{eqn:final_prob}
\end{multline}
\smallskip
At each timestep, we pick a derivation rule $R_k^n$ following ~\cref{eqn:final_prob} to expand the \frontier ($n_{f}^t$) in a depth-first, left-to-right fashion. When a terminal node is reached, it is recorded to be used in \tokenmodel and the decoder proceeds to next non-terminal. In~\Cref{eqn:final_prob}, $W_{tree}$ and $\mathbf{b}_{tree}$ are parameters that are jointly learned along with the LSTM parameters of the encoder and decoder.

\subsection{Token Generation Model (\tokenmodel)}
\label{sec:mtoken}

We now focus on generating a concrete code fragment $c$, \ie a sequence of tokens $(x_1, x_2, ...)$. 
For the edit task, the probability of an edited token $x^n_k$ depends not only on the tokens of the previous version ($x_1^p, ..., x_m^p$) but also on the previously generated tokens $x_1^n, ..., x_{k-1}^n$. The next token
$x^n_k$ also depends on the token type ($\theta$), which is generated by \treemodel. 
Thus, 
\smallskip
\begin{equation}
P(c_n|c_p, t_n) = \prod\limits_{k=1}^{m'}P(x_k^n | x_1^n, ..., x_{k-1}^n, \{x_1^p, ..., x_m^p\}, \theta_k^n)
    \label{eqn:token_prob}
\end{equation}
Here, $\theta_k^n$ is the node type corresponding to the generated terminal token $x_k^n$. 
Note that, the token generation model can be viewed as a conditional probabilistic translation model where token probabilities are conditioned not only on the context but also on the type of the token ($\theta_*^*$). 
Similar to \treemodel, we use an encoder-decoder. The encoder encodes each token and corresponding type of the input sequence into a hidden representation with an LSTM (\cref{fig:motiv_token}). Then, for each token ($x_i^p$) in the previous version of the code, the corresponding hidden representation ($s_i^p$) is given by: $\vect{s}_i^p = f_{LSTM}(\vect{s}_{i-1}^p, enc([x_i^p, \theta_i^p]))$.
Here, $\theta_i^p$ is the terminal token type corresponding to the generated token $x_i^p$ and $enc()$ is a function that encodes the pair of $x_i^p, \theta_i^p$ to a (learnable) vector representation.

The decoder's initial state is the final state of the encoder. Then, it generates a probability distribution over tokens from the vocabulary. The internal state at time step $k$ of the token generation is $\vect{s}_k^n = f_{LSTM}(\vect{s}_{k-1}^n, enc(x_i^n, \theta_k^n), \vect{\xi}_k))$,
where $\vect{\xi}_k$ is the attention vector over the previous version of the code and is computed as in \Cref{eqn:context}.
Finally, the probability of the $k$th target token is computed as
\smallskip
\begin{multline}
    P(x_k^n | x_1^n, ..., x_{k-1}^n, \{x_1^p, ..., x_m^n\}, \theta_k^n)\\
    = softmax\left(W_{token}\cdot\vect{s}_k^n + \vect{b}_{token}+ mask(\theta_k^n)\right)  
    \label{eqn:masked_prob}
\end{multline}
Here, $W_{token}$ and $\vect{b}_{token}$ are parameters that are optimized along with all other model parameters. Since not all tokens are valid for all the token types, we apply a mask that deterministically filters out invalid candidates.
For example, a token type of {\tt boolean\_value}, can only be concretized into \javaKeyWord{true} or \javaKeyWord{false}. Since the language grammar provides this information, we create a  mask ($mask(\theta_k^n)$) that returns a $-\infty$ value for masked entries and zero otherwise. Similarly, not all variable, method names, type names are valid at every position. We refine the mask based on the variables, method names and type names extracted from the scope of the change. 
{In the case of method, type and variable names, \tool allows \tokenmodel to generate a special {\tt <unknown>} token. However, the {\tt <unknown>} token is then replaced by the source token that has the highest attention probability (\ie the highest component of $\vect{\xi}_k$), a common technique in NLP. The mask restricts the search domain for tokens. However, in case to variable, type, and method name \tokenmodel can only generate whatever token available to it in the vocabulary (through masking) and whatever tokens are available in input code (through copying).}

\section{Implementation}
\label{sec:approach}

Our tree-based translation model is implemented as an edit recommendation tool, \tool. \tool learns source code changes from a dataset of patches. 
Then, given a code fragment to edit, \tool predicts potential changes that are likely to take place in a similar context. We implement \tool extending OpenNMT~\cite{2017opennmt} based on PyTorch. 
We now discuss \tool's implementation in details. 

\smallskip
\noindent
\textbf{Patch Pre-processing.} We represent the patches in a parse tree format and extract necessary information (\eg  grammar rules, tokens, and token-types) from them. 

\textit{Parse Tree Representation.}
As a first step of the training process, \tool takes a dataset of patches as input and parses them.
\tool works at method granularity. 
For a method patch $\Delta m$, \tool takes the two versions of $m$: \mold and \mnew. 
Using GumTree, a tree-based code differencing tool~\cite{falleri2014fine}, it identifies the edited AST nodes. The edit operations are represented as insertion, deletion, and update of nodes \wrt \mold. 
For example, in~\Cref{fig:motiv_pt}, \Red{red} nodes are identified as deleted nodes and~\Green{green} nodes are marked as added nodes. 
\tool then selects the minimal subtree of each AST that captures all the edited nodes. 
If the size of the tree exceeds a maximum size of $max\_change\_size$, we do not consider the patch. 
\tool also collects the edit context by including the nodes that connect the root of the method to the root of the changed tree. \tool expands the considered context until the context exceed a tree size threshold ($max\_tree\_size$).
During this process, \tool excludes changes in comments and literals. 
Finally, for each edit pair, \tool extracts a pair $(AST_{p}, AST_{n})$ where $AST_{p}$ is the original AST where a change was applied, and $AST_{n}$ is the AST after the changes.
\tool then converts the ASTs to their parse tree representation such that each token corresponds to a terminal node. 
Thus, a patch is represented as the pair of parse trees (\told, \tnew).

\textit{Information Extraction.}
\tool extracts grammar rules, tokens and token types from \told and \tnew.
To extract the rule sequence, \tool traverses the tree in a depth-first pre-order way. 
From \told, \tool records the rule sequence $(R_1^p,\: ...,\: R_\tau^p)$ and from \tnew, \tool gets $(R_1^n,\: ...,\: R_{\tau'}^n)$ (\Cref{fig:motiv_rules}). 
\tool then traverses the parse trees in a pre-order fashion to get the augmented token sequences, \ie tokens along with their terminal node types: $(x_*^p, \theta_*^p)$ from \told and $(x_*^n, \theta_*^n)$ from \tnew. 
\tool traverses the trees in a left-most depth-first fashion. 
When a terminal node is visited, the 
corresponding augmented token $(x_*^*, \theta_*^*)$ is recorded.


\noindent
\textbf{Model Training.}
We train the tree translation model (\treemodel) and token generation model (\tokenmodel) to optimize \Cref{eqn:tree_prob} and \Cref{eqn:token_prob} respectively using the cross-entropy loss as the objective function. Note that the losses of the two models are independent and thus we train each model separately. 
In our preliminary experiment, we found that the quality of the generated code is not entirely correlated to the loss. To mitigate this, we used top-1 accuracy to validate our model.
We train the model for a fixed amount of $n_{epoch}$ epochs using early stopping (with patience of $valid_{patience}$) on the top-1 suggestion accuracy on the validation data. 
We use stochastic gradient descent to optimize the model. 

\noindent
\textbf{Model Testing.}
To test the model and generate changes, we use beam-search~\cite{reddy1977speech} to produce the suggestions from \treemodel and \tokenmodel. 
{First given a rule sequence from the previous version of the tree, \tool generates \ktree rule sequences. \tool subsequently use these rule sequence to build the actual AST. While building the tree from the rule sequence, \tool ignores the sequence if the rule sequence is infeasible (\ie, head of the rule does not match the {\tt frontier\_node}, $n_f^t$). Combining the beam search in rule sequence and the tree building procedure, \tool generate different trees reflecting different structural changes.} Then for each tree, \tool generates \ktoken different concrete code. Thus, \tool generates \ktree $\cdot$ \ktoken code fragments. We sort them based on their probability, \ie $log(P(c_n|c_p, t_p)) = log( P(c_n|c_p, t_n) \cdot P(t_n|t_p))$.
From the sorted list of generated code, we pick the top $K$ suggestions. 

\begin{table*}[!t]
    \footnotesize
    \caption{{\small \textbf{Summary of datasets  used to evaluate \tool.}
    \vspace{-2mm}
    }}
    \label{tab:dataset_summary}
    \centering
        \begin{tabular}{l|r|r|r|r|r|r|r|r}
            \hlineB{2}
            \multirow{2}{*}{Dataset}  & \multirow{2}{*}{\# Projects}  & \#  Train & \# Validtion & \# Test & \multicolumn{2}{c|}{\# Tokens} & \multicolumn{2}{c}{\# Nodes} \bigstrut\\
              &  & Examples & Examples & Examples & Max & Average & Max & Average \bigstrut\\
           \hlineB{2}
            \dataname & 48 & 24072 & 3258 & 5143 & 38 & 15 & 47 & 20 \bigstrut\\
             \icsedata ~\cite{chen2018sequencer} & 3 & 4320 & 613 & 613 & 34 & 17 &  47 & 23\bigstrut\\
            \defj-data~\cite{just2014defects4j} & 6 & 22060 & 2537 & 117 & 35 & 16 &  43 & 21\bigstrut\\
            
             \hlineB{2}
        \end{tabular}
    \vspace{-7mm}
\end{table*}

\section{Experimental Design}
\label{sec:experiment}
\label{subsec:study-subject}

We evaluate \tool for three different types of changes that often appear in practice: 
(i) code change in the wild, (ii) pull request edits, and (iii) bug repair. For each task, we train and evaluate \tool on different datasets. \Cref{tab:dataset_summary} provides detailed statistics of the datasets we used. 

 
\smallskip
\noindent
(i) \textit{Code Change Task.} 
{We collected a large scale real code change dataset (\dataname) from 48 open-source projects from GitHub. These projects also appear in TravisTorrent~\cite{msr17challenge} and have 
aat least 50 commits in Java files. These project are  excludes any forked project, toy project, or unpopular projects (all the projects have at least 10 watchers in Github). Moreover, these projects are big and organized enough that they use Travis Continuous integration system for maintainability. For each project, we collected the revision history of the main branch. For each commit, we record the code before and after the commit for all Java files that are affected. In total we found java 241,976 file pairs.} 
{We then use GumTree~\cite{falleri2014fine} tool to locate the change in the file and check whether the changes are inside a method in the corresponding files. Most of the changes that are outside a method in a java class are changes related to import statements and changes related to constant value. We consider those out of \tool's scope. We further remove any pairs, where the change is only in literals and constants. Excluding such method pairs, we got 182,952 method pairs. We set $max\_change\_size=10$ and $max\_tree\_size=20$. With this data collection hyper-parameter settings, we collected 44,382 patches.}

{We divide every project based on their chronology. From every project, we divide earliest 70\% patches into train set, next 10\% into validation set and rest 20\% into test set based on the project chronology.} 
{We removed any exact test and validation examples from the training set. We also removed intra-set duplicates. After removing such duplicate patches, we ended up with 32,473 patches in total, which are then used to train and evaluate \tool.}

\smallskip
\noindent
(ii) \textit{Pull Request Task.} For this task, we use \icsedata, provided by Tufano \etal~\cite{tufano2019learning} which contains source code changes from merged pull requests from three projects from Gerrit~\cite{gerrit}. Their dataset contains 21774 method pairs in total. 
{Similar to the \dataname, we set $max\_change\_size=10$ and $max\_tree\_size=20$ to extract examples that are in \tool's scope from this dataset. We extracted 5546 examples patch pair.}

\smallskip
\noindent
(iii) \textit{Bug Repair Task.} For this task, we evaluate \tool on \defj~\cite{just2014defects4j} bug-fix patches. We extract 117 patch pairs across 80 bug ids in Defect4j dataset with the same  $max\_change\_size=10$ and $max\_tree\_size=20$ limit as Code Change Task. These are the bugs that are in \tool's scope. To train \tool for this task, we create a dataset of code changes from six projects repositories in \defj dataset containing 24597 patches. We remove the test commits from the training dataset. 

\subsection{Evaluation Metric}
\label{subsec:evaluation-metric}
To evaluate \tool, we measure for a given code fragment, how accurately \tool generates patches. We consider \tool to correctly generate a patch if it exactly matches the developer produced patches. \tool produces the top $K$ patches and we compute \tool's accuracy by counting how many patches are correctly generated in top $K$. 
Note that this metric is stricter than semantic equivalence.

For the bug fix task,~\tool takes a buggy line as input and generates the corresponding patched code. We consider a bug to be fixed if we find a patch that passes all the test cases. We also manually investigate the patches to check the similarity with the developer provided patches.

\subsection{Baseline}
\label{subsec:baseline}

\noindent
We consider several baselines to evaluate \tool's performance. Our first baseline in a vanilla LSTM based \sts model with attention mechanism~\cite{bahdanau2014neural}. Results of this baseline indicate different drawbacks of considering raw code as a sequence of token.


{The second baseline, we consider, is proposed by Tufano~\etal~\cite{tufano2018nmt_bug_fix}. For a given code snippet (previous version), Tufano~\etal first abstracted the identifier names and stored the mapping between original and abstract identifiers in a symbol table. The resultant abstracted code (obtained by substituting the raw identifiers with abstract names) is then translated by an NMT model. After translation, they concretize the abstract code using the information stored in the symbol table.  The patches where the abstract symbols predicted by the model are not found in the symbol table remian undecidable. Such patches, although can be useful to guide developers similar to our \treemodel, cannot be automatically concretized, and thus, we do not count them as fully correct patches.}

Both the vanilla \sts and Tufano~\etal's model consider the before version of the code as input. Recently, SequenceR~\cite{chen2018sequencer} proposed way to represent additional context to help the model generate concrete code. We design such a baseline, where we add additional context to $c_P$. Following SequenceR, we add copy attention, where the model learns to copy from the contexed code.

{To understand the tree encoding mechanism, we used several tree encoders. First method we considered is similar to DeepCom~\cite{hu2018deep}, where the AST is represented as a sequential representation called Structure Based Traversal (SBT). Second tree encoding method we consider is similar to Code2Seq, where code AST is represented by a set of paths in the tree. While these tree encoding methods are used for generating Code comment, we leverage these encoding methods for code change prediction. We design a \sts method with DeepCom encoder (Tree2Seq), and Code2Seq encoder. We also enable the copy attention in both of these baselines.}



{We also did some experiment on Kommrusch~\etal~\cite{kommrusch2020equivalence}'s proposed Graph2Seq network, which they provided implementation as an add on in OpenNMT-Py framework\footnote{\url{https://opennmt.net/OpenNMT-py/ggnn.html}}. However, their implementation is assumed upon a very restricted vocabulary size. To understand better, we modified their code to fit 
into our context, but neither their original model, nor our 
modified model works in program change task. When we increase the vocabulary size in their version of the model, it does not scale and quickly exhaust all the memory in the machine (we tried their model on a machine with 4 Nvidia TitanX GPU and 256 GB RAM). When we tested our modification of their code, the model does not converge even after 3 days of training, eventually resulting in 0 correct predictions.}

The basic premise of~\tool is based on the fact that code changes are repetitive. Thus, another obvious baseline is to see how \tool performs \wrt to code-clone based edit recommendation tool~\cite{ray2014uniqueness}. In particular, given a previous version ($c_p$) of an edit, we search for the closest $k$ code fragments using similar bag-of-words at the token level similar to Sajnani \etal~\cite{sajnani2016sourcerercc}. In our training data of code edits, this step searches in the previous code versions and use the corresponding code fragments of the next version as suggested changes.

\noindent
\textit{Bug-fixing baselines}: For the bug fix task, we compare \tool's performance with two different baselines. Our first baseline is SequenceR~\cite{chen2018sequencer}, we compare with the results they reported. We also compare our result with other non-NMT based program repair systems.

\section{Results}
\label{sec:findings}

We evaluate \tool's performances to generate concrete patches \wrt generic edits (RQ1) and bug fixes (RQ3).
In RQ2, we present an ablation study to evaluate our design choices.

\RQ{1}{\rqa}

\begin{table*}[!htb]
  \centering
  \footnotesize
  \caption{\textbf{\small{Performance  of  \tool suggesting concrete patches. For Token Based models, predominant source of information in the code are token sequences. For Tree Based models information source is code AST. For IR based method, information retrieval model is used on code.}}
  \vspace{-2mm}
  }
  \label{tab:concrete}%
    \begin{tabular}{l|l|c|c|c|c|c|c}
\hlineB{2}
\multicolumn{2}{c|}{\multirow{3}{*}{\textbf{Method}}} & \multicolumn{3}{c|}{\textbf{Code Change Data}} & \multicolumn{3}{c}{\textbf{Pull Request Data}}  \bigstrut\\ \cline{3-8} 
\multicolumn{2}{l|}{}  & \multicolumn{3}{c|}{\textbf{Number of examples : 5143}} & \multicolumn{3}{c}{\textbf{Number of examples : 613}}                           \bigstrut\\ \cline{3-8} 
\multicolumn{2}{l|}{}   &\textbf{ Top-1}     & \textbf{Top-2}     & \textbf{Top-5}      & \textbf{Top-1}     & \textbf{Top-2}    & \textbf{Top-5}                  \bigstrut\\ \hlineB{2}
\multirow{3}{*}{\textbf{Token Based}}  & \sts       & 107 (2.08\%)          & 149 (2.9\%)            & 194 (3.77\%)           & 45 (7.34\%)           & 55 (8.97\%)            & 69 (11.26\%)           \bigstrut\\ \cline{2-8} 
                              & Tufano~\etal & 175 (3.40\%)          & 238 (4.63\%))          & 338 (6.57\%)           & \textbf{81 (13.21\%)} & 104 (16.97\%)           & 145 (23.65\%)          \bigstrut\\ \cline{2-8} 
                              & SequenceR     & \textbf{282 (5.48\%)}          & 398 (7.74\%)           & 502 (9.76\%)          & 39 (6.36\%)           & \textbf{137 (22.35\%)} & 162 (26.43\%)          \bigstrut\\ \hlineB{2}
\multirow{3}{*}{\textbf{Tree Based}}   & Tree2Seq      & {147 (2.86\%)} & 355 (6.9\%)           & 568 (11.04\%)          & 39 (6.36\%)           & 89 (14.52\%)           & 144 (23.49\%)          \bigstrut\\ \cline{2-8} 
                              & Code2seq      & 58 (1.12\%)           & 82 (1.59\%)            & 117 (2.27\%)           & 4 (0.65\%)            & 7 (1.14\%)             & 10 (1.63\%)            \bigstrut\\ \cline{2-8} 
                              & \tool         & 201 (3.91\%)          & \textbf{571 (11.10\%)} & \textbf{820 (15.94\%)} & 57 (9.3\%)            & 134 (21.86\%)          & \textbf{177 (28.87\%)} \bigstrut\\ \hlineB{2}
\textbf{IR based }                     & \ir           & 40 (0.77\%)           & 49 (0.95\%)            & 61 (1.18\%)            & 8 (1.30\%)            & 8 (1.30\%)             & 9 (1.46\%)             \bigstrut\\ \hlineB{2}
\end{tabular}
    \vspace{-6mm}
\end{table*}%

To answer this RQ, we evaluate \tool's accuracy \wrt the evaluation dataset containing concrete patches. Table~\ref{tab:concrete} shows the results: for \dataname, \tool can successfully generate $201$ ($3.91$\%), $571$ ($11.10$\%), and $820$ ($15.94$\%) patches at top 1, 2, and 5 respectively.
In contrast, at top 1, SequenceR generates $282$ ($5.48\%$) correct patches, and performs the best among all the methods. While SequenceR outperforms \tool in top 1, \tool outperforms SequenceR with significant margin at top 2 and top 5. 

In \icsedata, \tool generates $57$ ($9.3\%$), $134$ ($21.86\%$), and $177$ ($28.87\%$) correct patches at top 1, 2, and 5 respectively. At top 1, Tufano~\etal's~\cite{tufano2018nmt_bug_fix} method produces 81 patches. At top 2, \tool produces $134$ ($21.86\%$) patches, which is comparable with SequenceR's result $137$ ($22.35\%$). At top 5, \tool outperforms all the other baselines achieving 9.2\% gain over SequenceR baseline.

{The main advantage point of \tool is that, since it considers the structural changes separate from the token changes, it can learn the structural change pattern well instead of being dominated by learning the code token changes. However, being a two stage process, \tool has two different hinge point for failure. If \treemodel does not generate the correct tree, no matter how good \tokenmodel performs, \tool is unable to generate correct patch. We conjecture that, this is the reason for \tool's failure at top 1. }

Among the baselines we compared here, SequenceR, and Tree2Seq takes the advantage of copy attention. Tufano~\etal's model takes the advantage of reduced vocabulary through identifier abstraction. Tufano~\etal's model works best when the code is mostly self contained \ie, when there is always a mapping from abstract identifier to concrete identifier in the symbol table, their method can take full advantage of the vocabulary reduction. to When the input code is mostly self contained (\ie, most of the tokens necessary to generate a patch are inside the input code $c_p$ or appears within the context limit in as presented to SequenceR). 

{In code change task with NMT, the deterministic positions of code tokens are important to put attention over different parts of code. While Code2Seq~\cite{alon2018code2seq}'s representation of input code as random collection of paths in the code showed success for code comprehension, it does not generalize well for code change due to the stochastic nature of the input. Additionally, copy attention cannot be trivially applied to Code2Seq since, like attention, copy also rely on the defined positions of tokens in the input.}

While \tool replaces any {\tt <unknown>} prediction by the token with highest attention score (see~\cref{sec:mtoken}), unlike SequenceR, \tool does not learn to copy. The rationale is, unlike SequenceR, \tool's input code ($c_p$) is not enhanced by the context. Instead, we present the context to the \tool through the token mask (see~\cref{eqn:masked_prob}). If we enable copy attention, \tool becomes highly biased by the tokens that are inside $c_p$. 

\begin{figure}[!htb]
    \centering
    \scriptsize
    \subfigure[\textbf{\dataname}]{
    \includegraphics[width=0.45\textwidth]{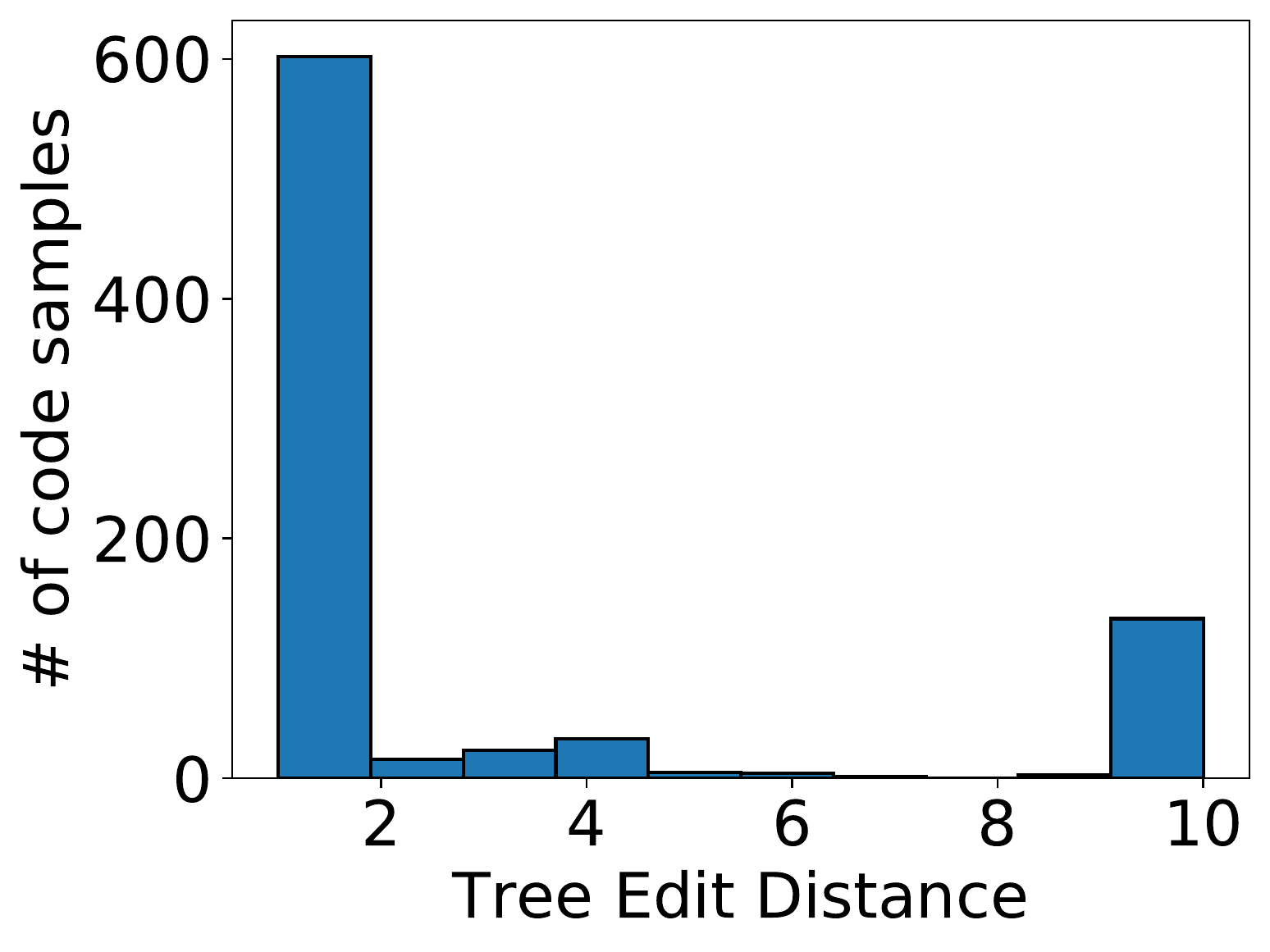}
    \label{fig:ed_hist_codit}
    }
    \subfigure[\textbf{\icsedata}]{
    \includegraphics[width=0.45\textwidth]{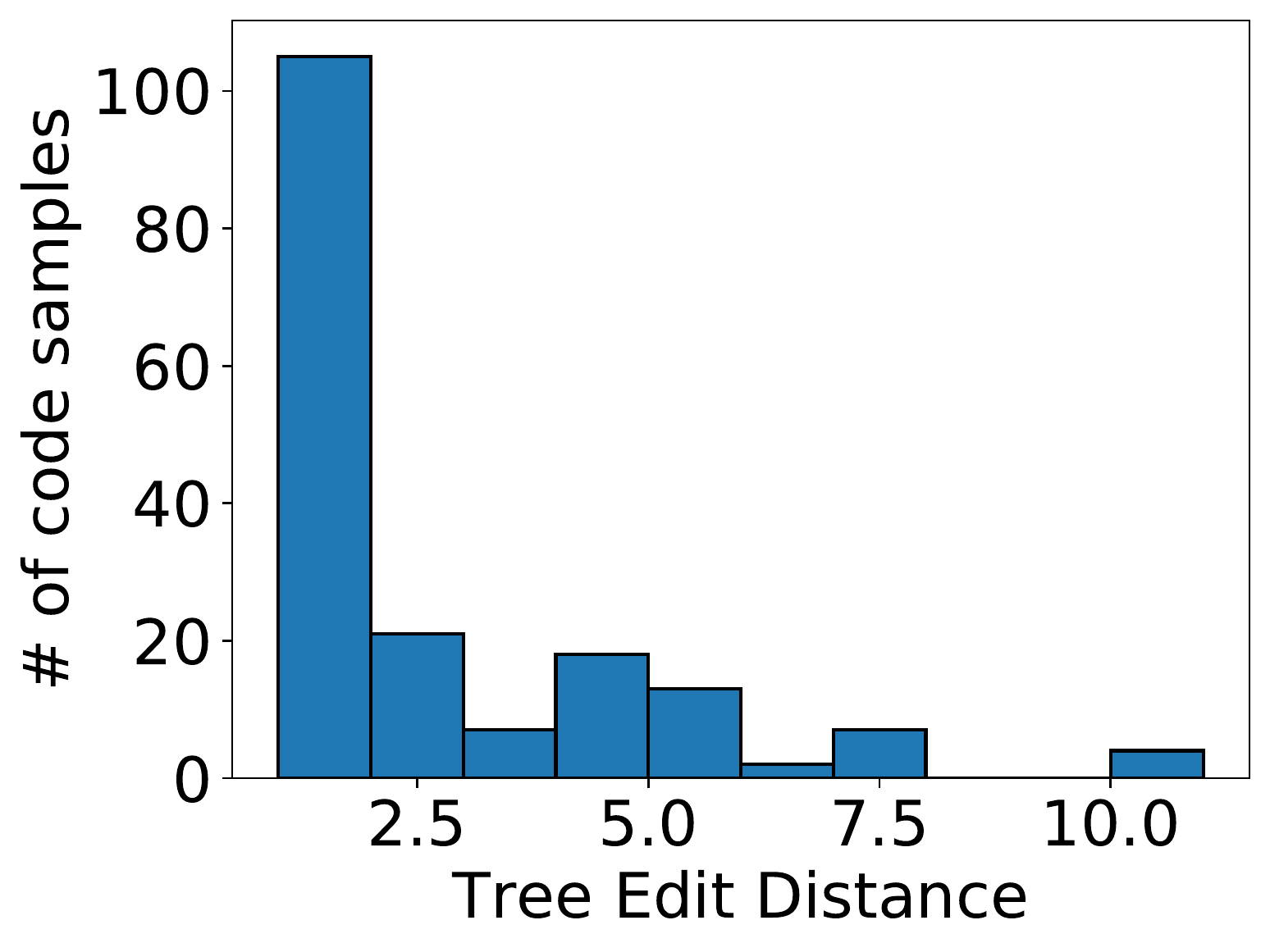}
    \label{fig:ed_hist_icse}
    }
    \caption{\small{\textbf{Patch size (Tree Edit-distance) histogram of correctly generated patches in different datasets.
    }}}
    \label{fig:ed_hist}
\end{figure}

\begin{table}[t]
\caption{\textbf{\small{Examples of different types of concrete patches generated by \tool}}
}
 \label{tab:concrete_examples}
\centering
\small
\setlength{\tabcolsep}{2pt}
\begin{tabular}{p{.9\columnwidth}}
\toprule
\textbf{Example 1.~API Change} \\ 
\begin{lstlisting}
return f.@\Red{\sout{createJsonParser}} \Green{\underline{createParser}}@(...)
\end{lstlisting} \\
\midrule

\textbf{Example 2.~Type Change} \\ 
\begin{lstlisting}
void appendTo(@\Red{\sout{StringBuffer}} \Green{\underline{StringBuilder}}@ buffer) 
\end{lstlisting} \\
\midrule
\textbf{Example 3.~Parameter: Add/Delete Method Parameter} \\ 
\begin{lstlisting}
1. testDataPath(false , true , true , true@\Green{\underline{, false} }@);  
2. assertNotificationEnqueued(map ,key ,value @\Red{\sout{,hash}}@) 
\end{lstlisting}\\
\midrule

\textbf{Example 4.~Refactoring: Modify Method Parameters Name} \\
\begin{lstlisting}
void visit(JSession @\Red{\sout{x}} \Green{\underline{session}}@ , ...) throws Exception 
{ 
    visit (((JNode) (@\Red{\sout{x}} \Green{\underline{session}}@)), ...); 
}
\end{lstlisting} \\
\midrule

\textbf{Example 5.~Statement: Add Statement} \\ 
\begin{lstlisting}
{... 
    interruptenator.shutdown(); 
    @\Green{\underline{Thread.interrupted(); }}@
}
\end{lstlisting} \\
\midrule

\textbf{Example 6.~Inheritance: Abstracting a Method} \\ 
\begin{lstlisting}
public @\Green{\underline{abstract}}@ void removeSessionCookies (...)
 @\Red{\sout{\{}}@ 
     @\Red{\sout{throw new android...MustOverrideException(); }}@
 @\Red{\sout{\}}}@
\end{lstlisting} \\

\midrule

\textbf{Example 7.~Exception Change: Add Try Block} \\ 
\begin{lstlisting}
 public void copyFrom( java.lang.Object arr){ 
 @\Green{+} @    @\Green{try\{}@ 
 	android.os.Trace.traceBegin (...); 
 @\Green{+} @    @\Green{finally\{}@ 
 	android.os.Trace.traceEnd(...); 
 @\Green{+} @    @\Green{\}}@
 }		
\end{lstlisting} \\

 \midrule

\textbf{Example 8.~Other: Delete Unreferenced Variable} \\ 
\begin{lstlisting}
 public void testConstructor2NPE(){ 
 ...
 @\Red{-\sout{AtomicIntegerArray aa =}}@ new AtomicIntegerArray(a);
 shouldThrow () ; 
 ...
 }
\end{lstlisting} \\
\midrule

\textbf{Example 9.~Final: Make Method Parameter Immutable} \\ 
\begin{lstlisting}
public String print(@\Green{\underline{final}}@ ReadableInstant inst){...}
\end{lstlisting} \\
\midrule

\textbf{Example 10.~Access: Modify Access Type} \\ 
\begin{lstlisting}
 @\Red{\sout{public}} \Green{\underline{protected}}@ AbstractInstant () { super(); }
\end{lstlisting} \\
\bottomrule

\end{tabular}
\begin{flushleft}
    \scriptsize{Every cell shows an example of correctly suggested patches by \tool. Top line is the patch category, followed by the actual patch. In the patch, {\tt \Red{\sout{Red}}} tokens/lines are deleted and {\tt \Green{\underline{Green}}} tokens/lines are added.}
\end{flushleft}
\label{examples}
\end{table}

\begin{table}[h]
\scriptsize
\centering
\caption{\small{\textbf{Examples of \tool's ability to generalize  in different use cases. {\tt Exception}, {\tt Error}, {\tt RuntimeException} are modified to {\tt EOFException} under different context.
\vspace{-2mm}}}}
\label{tab:similar_examples}
\begin{tabular}{p{.90\columnwidth}}
\toprule
\begin{lstlisting}[label={lst:ex1}]
Ex:1. return @\Red{\sout{Error}} \Green{\underline{EOFException}}@.class ; 
\end{lstlisting}\\
\begin{lstlisting}[label={lst:ex2}]
Ex:2. return @\Red{\sout{Exception}} \Green{\underline{EOFException}}@.class ; 
\end{lstlisting}\\
\begin{lstlisting}[label={lst:ex3}]
Ex:3. return @\Red{\sout{RuntimeException}} \Green{\underline{EOFException}}@.class ; 
\end{lstlisting}\\
\begin{lstlisting}[label={lst:ex4}]
Ex:4. return new @\Red{\sout{Error}} \Green{\underline{EOFException}}@(msg) ; 
\end{lstlisting}\\
\begin{lstlisting}[label={lst:ex5}]
Ex:5. return new @\Red{\sout{Exception}} \Green{\underline{EOFException}}@(msg); 
\end{lstlisting}\\
\begin{lstlisting}[label={lst:ex6}]
Ex:6. return new @\Red{\sout{RuntimeException}} \Green{\underline{EOFException}}@(msg); 
\end{lstlisting}\\
\bottomrule
\end{tabular}
\end{table}

{Note that, \emph{``vocabulary explosion''} still remains an open problem for code generation. Neither \tool nor any other baselines we discussed here solve this problem.  \tool presents a way to learn the structural change and restricts the search domain for token names through a mask. For instance,  where \tokenmodel needs to generate a token of \emph{primitive data type} (\tokenmodel knows the token type because \treemodel already generated a tree with terminal node types), it can restrict the search over the primitive types only. While it is expected that the decoder of an ideal \sts model would inherently learn appropriate tokens at appropriate positions as it implicitly learns code structure, in reality, because of its unrestricted search space, they tend to mispredict more tokens than \tool. }


In general, \tool along with all the baselines perform better when generating small patches. For example, a large majority of the correctly generated patches have size of one (\ie $\Delta_t$ = 1, the tree distance between \told and \tnew~\cite{masek1980faster}). However, a non-trivial number of larger patches are also correctly generated. \Cref{fig:ed_hist} shows the histogram of the size of correctly predicted patches. 
For example, there are 202 and 51 correct patches with $\Delta_{t} \geq 3$ generated by \tool for \dataname and \icsedata respectively.  

For qualitative evaluation, we show some non-trivial patches that \tool can successfully generate. \tool can learn a wide range of patch patterns. \Cref{examples} shows few examples of different category of patches that \tool can generate. \tool also shows promise in generating non-trivial structure based changes. Consider Example 4 where {\tt x} is renamed to {\tt session} both the formal parameter and the usage in the body. Note that, since \tool uses a tree-based model it is good at capturing long-distance dependencies allowing the token-level model to focus on predicting tokens, \eg such that it can rename the same variable similarly. Another interesting example where \tool can successfully generate patches is shown in example 6, where \tool does not only add the \javaKeyWord{abstract} keyword in the method signature, but also removes the body. Since \tool is aware of code syntax, it learns that method declarations with an \javaKeyWord{abstract} keyword have a high probability of an empty method body. 

\Cref{tab:similar_examples} shows some additional examples where \tool can successfully generate patches. In these examples, different exception/error types (\ie~{\tt Exception}, {\tt Error}, {\tt RuntimeException}) are changed to {\tt EOFException} although their usage differs. In the first three examples {\tt EOFException} is used as a class reference, while for the others {\tt EOFException} is used to initialize an object. These examples also illustrate \tool's ability to generalize to different contexts and use-cases.  Other structural transformation that \tool include, but not limited to, include scoping (example 7 in~\cref{examples}), adding/deleting method parameters (example 3 in~\cref{examples}), changing method/variable access modifiers (example 9, 10 in~\cref{examples}), etc. 

\RS{1}{\tool suggests  15.94\% correct patches for \dataname and 28.87\% for \icsedata within top 5 and outperforms best baseline by 44.37\% and 9.26\% respectively. }

Next, we evaluate \tool's sub-components.\\ 
\RQ{2}{\rqb}
This RQ is essentially an ablation study where we investigate  in three parts: 
(i)  the token generation model (\tokenmodel), 
(ii) the tree translation model (\treemodel), and
(iii) evaluate the combined model, \ie \tool, under different design choices.
We further show the ablation study on the data collection hyper-parameters (\ie, $max\_change\_size$, and $max\_tree\_size$) and investigate the cross project generalization.  

\noindent
\textbf{Evaluating token generation model.}
Here we compare \tokenmodel with the baseline SequenceR model. Note that \tokenmodel generates a token given its structure. Thus, for evaluating the standalone token generation model in \tool's framework, we assume that the true structure is given (emulating a scenario where a developer knows the kind of structural change they want to apply). ~\Cref{tab:gold_tree} presents the results. \begin{table}[!htb]
    \footnotesize
    \centering
    \caption{\textbf{\small Correct patches generated by the standalone token generation model when the true tree structure is known.
    }}
    \label{tab:gold_tree}
    \begin{tabular}{c|c|c}
       \hlineB{2}
         \multirow{2}{*}{Dataset} &\multicolumn{2}{c}{ Total Correct Patches}\bigstrut\\
         \cline{2-3}
          & SequenceR & standalone \tokenmodel \bigstrut\\
         \hlineB{2}
         \dataname & 502 (9.76\%) & 2035 (39.57\%)\bigstrut\\
         \icsedata & 162 (26.43\%) & 378 (61.66\%)\bigstrut\\
         \hlineB{2}
    \end{tabular}
    \vspace{-3mm}
\end{table}

While the baseline \sts with copy-attention (SequenceR) generates 9.76\% (502 out of 5143) and 26.43\% (162 out of 613) correct patches for \dataname and \icsedata respectively at top 5, \Cref{tab:gold_tree} shows that if the change structure  (\ie \tnew) is known, the standalone \tokenmodel model of \tool can generate 39.57\% (2035 out of 6832) and 61.66\% (378 out of 613) for \dataname and \icsedata respectively. 
This result empirically shows that if the tree structure is known, NMT-based code change prediction significantly improves. In fact, this observation led us build \tool as a two-stage model.


\begin{table}[!htb]
    \centering
    \footnotesize
    \caption{\textbf{\small{\treemodel top-5 performance for different settings.
    }}
    }
    \label{tab:tree_prediction}
    \begin{tabular}{c|c|c}
    \hlineB{2}
         \multirow{2}{*}{Dataset} &\multicolumn{2}{c}{ \# Correct Abstract Patches$^*$}\bigstrut\\
         \cline{2-3}
          & Tufano~\etal  & \tool\bigstrut\\
         \hlineB{2}
         \dataname & 1983 / 5143 (38.56\%) & 2920 / 5143 (56.78\%)\bigstrut\\
         \icsedata & 241 / 613 (39.31\%) & 342 / 613 (55.79\%)\bigstrut\\
         \hlineB{2}
    \end{tabular}

    {\scriptsize * Each cell represents correctly predicted patches / total patches (percentage of correct patch) in the corresponding setting.}
\end{table}

\noindent
\textbf{Evaluating tree translation model.}
{Here we evaluate how accurately \treemodel predicts the structure of a change --- shown in \Cref{tab:tree_prediction}. \treemodel can predict 56.78\% and 55.79\% of the structural changes in \dataname and \icsedata respectively. Note that, the outputs that are generated by \treemodel are not concrete code, rather structural abstractions. In contrast, Tufano~\etal's~\cite{tufano2018nmt_bug_fix} predicts 38.56\% and 39.31\% correct patches in \dataname and \icsedata respectively. Note that, their abstraction and our abstraction method is completely different. In their case, for some of the identifiers (those already exist in the symbol table), they have a deterministic way to concretize the code. In our case, \tokenmodel in \tool augments \treemodel by providing a stochastic way to concretize every identifier by using NMT. }



Note that, not all patches contain structural changes (\eg when a single token, such as a method name, is changed).  For example, 3050 test patches of \dataname, and 225 test patches of \icsedata do not have structural changes. When we use these patches to train \treemodel, we essentially train the model to sometimes copy the input to the output and rewarding the loss function for predicting no transformation.
Thus, to report the capability of \treemodel to predict structural changes, we also train a separate version of \treemodel using only the training patches with at least 1 node differing between \tnew and \told.  We also remove examples with no structural changes from the test set. This is our filtered dataset. In the filtered dataset, \treemodel predicts 33.61\% and 30.73\% edited structures from \dataname and \icsedata respectively. This gives us an estimate of how well \treemodel can predict \emph{structural} changes.

\medskip
\noindent
\textbf{Evaluating \tool.} 
Having \treemodel and \tokenmodel evaluated separately, we will now evaluate our end-to-end combined model (\treemodel $+$ \tokenmodel) focusing on two aspects: (i) effect of attention-based copy mechanism, (ii) effect of beam size.

First, we evaluate contribution of \tool's attention-based copy mechanism as described in~\Cref{sec:mtoken}. \Cref{tab:replace_unk_summary} shows the results. Note that, unlike SequenceR, \tool is not trained for learning to copy. Our copy mechanism is an auxiliary step in the beam search that prohibits occurrence of {\tt <unknown>} token in the generated code. 

\begin{table}[t]
    \centering
    \footnotesize
    \caption{\small \textbf{\tool performance \wrt to the attention based copy mechanism @top-5 (\ktree=2, \ktoken=5).}
    Lower bound is without copy.
    The upper bound evaluates with oracle copying predictions for {\tt <unknown>}. 
    For \tool each {\tt <unknown>} token is replaced by the source token with the highest attention.
    \vspace{-1mm}
    }
    \label{tab:replace_unk_summary}
    \begin{tabular}{c|c|c|c}
    \toprule
                
         Dataset & lower bound & upper bound & \tool\\
         \midrule
         \dataname & 742 (14.42\%) & 898 (17.46\%) & 820 (15.94\%)\\
         \midrule
         \icsedata & 163 (26.59\%) &  191 (31.16\%) & 177 (28.87\%)\\
         \bottomrule
    \end{tabular}
     \vspace{-2mm}
\end{table}



\begin{figure}[!htb]
    \centering
    \includegraphics[width=0.7\linewidth]{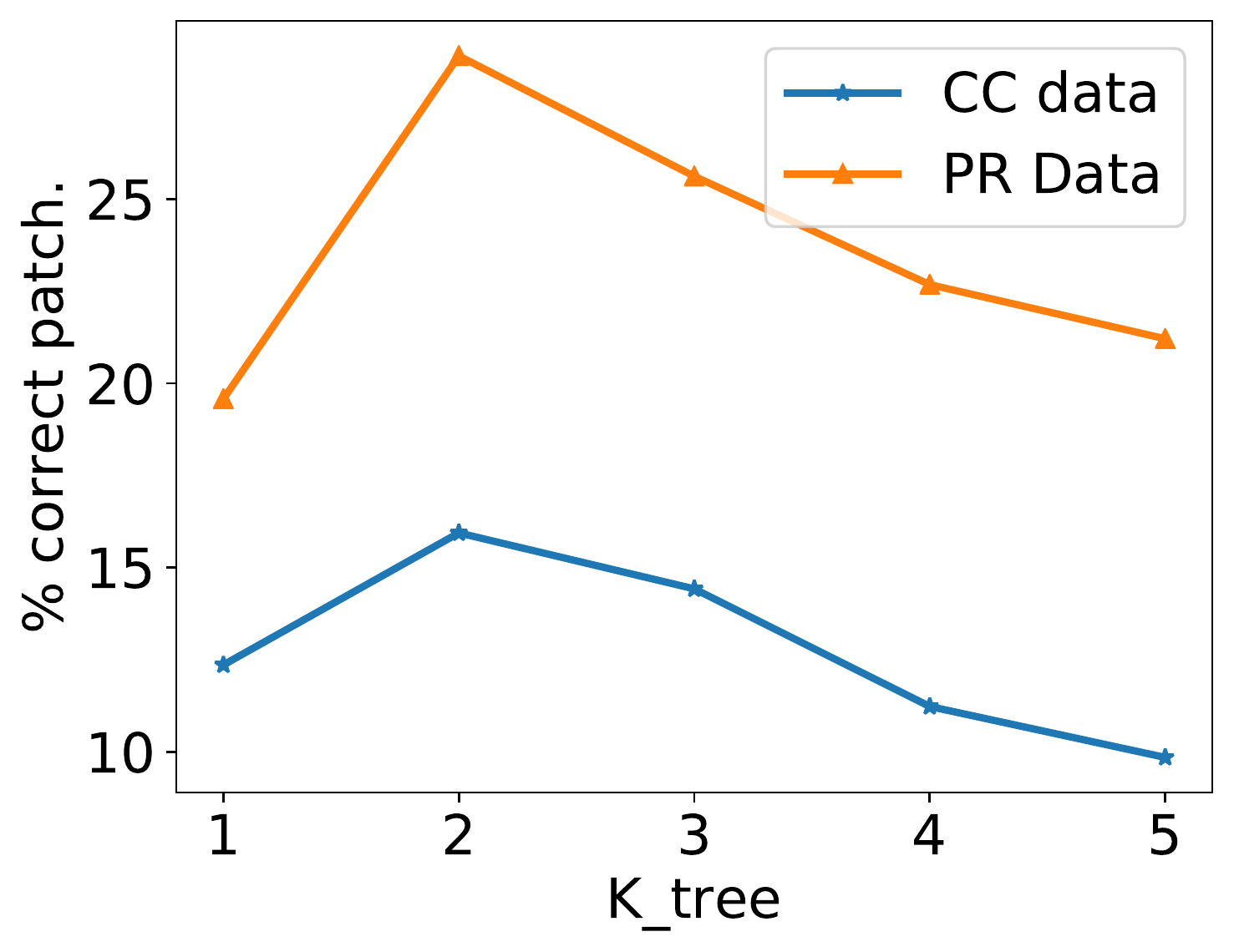}
    \caption{\small{\textbf{Performance of \tool @top-5 (\ktoken = 5) for different \ktree}}}
    \label{tab:k1k2}
\end{figure}

Recall that \tokenmodel generates a probability distribution over the vocabulary (\Cref{sec:mtoken}). Since the vocabulary is generated using the training data, any unseen tokens in the test patches are replaced by a special {\tt <unknown>} token. In our experiment, we found that a large number (about 3\% is \dataname and about 4\% is \icsedata) of patches contain {\tt <unknown>} tokens; this is undesirable since the generated code will not compile. When we do not replace {\tt <unknown>} tokens, \tool can predict {742 (14.42\%)}, and {163 (26.59\%)} correct patches in \dataname and \icsedata respectively. However, if all the {\tt <unknown>} tokens could be replaced perfectly with the intended token, \ie upper bound of the number of correct patches goes up to {898 (17.46\%)} and {191 (31.16\%)} correct patches in \dataname and \icsedata respectively. This shows the need for tackling the {\tt <unknown>} token problem. To solve this, we replace {\tt <unknown>} tokens predicted by \tokenmodel with the source token with the highest attention probability following~\Cref{sec:mtoken}. With this, \tool generates {820 (15.94\%)}, and {177 (28.87\%)} correct patches from \dataname and \icsedata respectively (\Cref{tab:replace_unk_summary}).

{Second, we test two configuration parameters related to the beam size, \ktree and \ktoken \ie the number of trees generated by \treemodel and number of concrete token sequences generated by \tokenmodel per tree (\Cref{sec:approach}). While \Cref{tab:concrete} shows the effect of different values of \ktoken (\eg, 1, 2, 5), here we investigate the effect of \ktree for a given \ktoken. \Cref{tab:k1k2} shows the parameter sensitivity of \ktree when \ktoken is set 5. Recall from \Cref{sec:approach}, \tool first generates \ktree number of trees, and then generates \ktoken number of code per tree. Among those \ktree$ * $\ktoken generated code, \tool chooses top \ktoken number of code to as final output. In both \dataname (CC data in \cref{tab:k1k2}), and \icsedata (PR data in \cref{tab:k1k2}), \tool performs best when \ktree = 2. When \ktree = 2, total generated code is 10, among which \tool chooses top 5. With increasing number of \ktree, \tool has to choose from increasing number of generated code, eventually hurting the performance of \tool.}

\begin{table}[!htb]
    \centering
    \footnotesize
    \caption{{\small \textbf{Ablation study of $max\_change\_size (max_c)$, and $max\_tree\_size (max_t)$.}}}
    \begin{tabular}{c|c|c|c|c|c}
    \hlineB{2}
    \textbf{\# Train} & \textbf{\# Valid} & \textbf{\# Test} & \textbf{$max_c$} &\textbf{ $max_t$} & \textbf{\% correct} \bigstrut\\
    \hlineB{2}
    3988 & 593 & 593 & 10   & 10 & 34.06 \bigstrut\\
    4320 & 613 & 613 & 10   & 20 & 32.63 \bigstrut\\
    4559 & 622 & 627 & 10   & 30 & 28.07 \bigstrut\\
    \hline
    7069 & 948 & 932 & 20   & 20 & 31.55 \bigstrut\\
    7340 & 993 & 948 & 20   & 30 & 29.32 \bigstrut\\
    \hline
    9361 & 1227 & 1213 & 30   & 30 & 24.48 \bigstrut\\
    \hlineB{2}
    \end{tabular}
    \label{tab:tree_size_ablation}
\end{table}

\begin{figure}[!htb]
    \centering
    \scriptsize
    \subfigure[\textbf{\dataname}]{
    \includegraphics[width=0.47\textwidth]{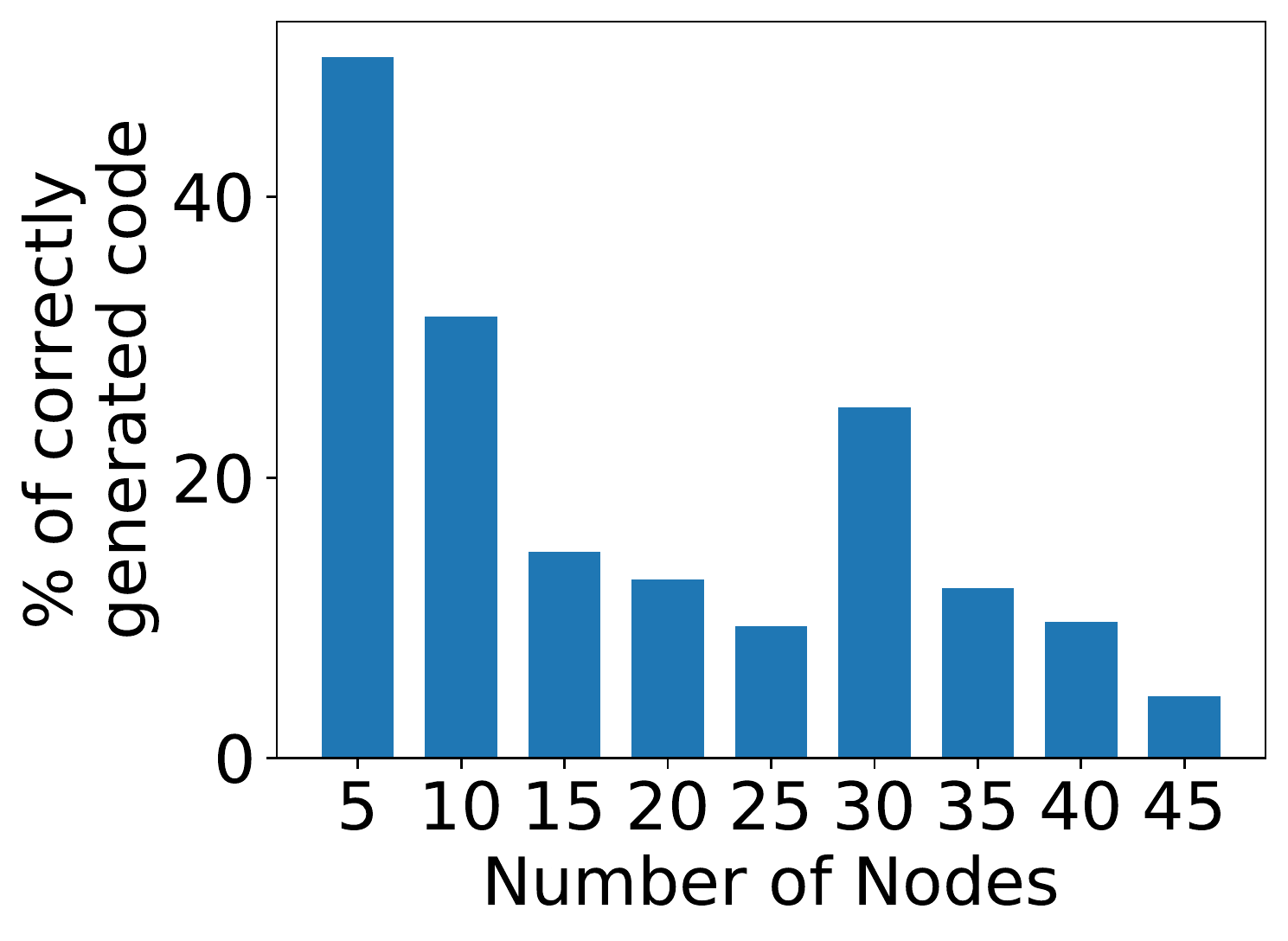}
    \label{fig:cc_tree_size}
    }
    \subfigure[\textbf{\icsedata}]{
    \includegraphics[width=0.47\textwidth]{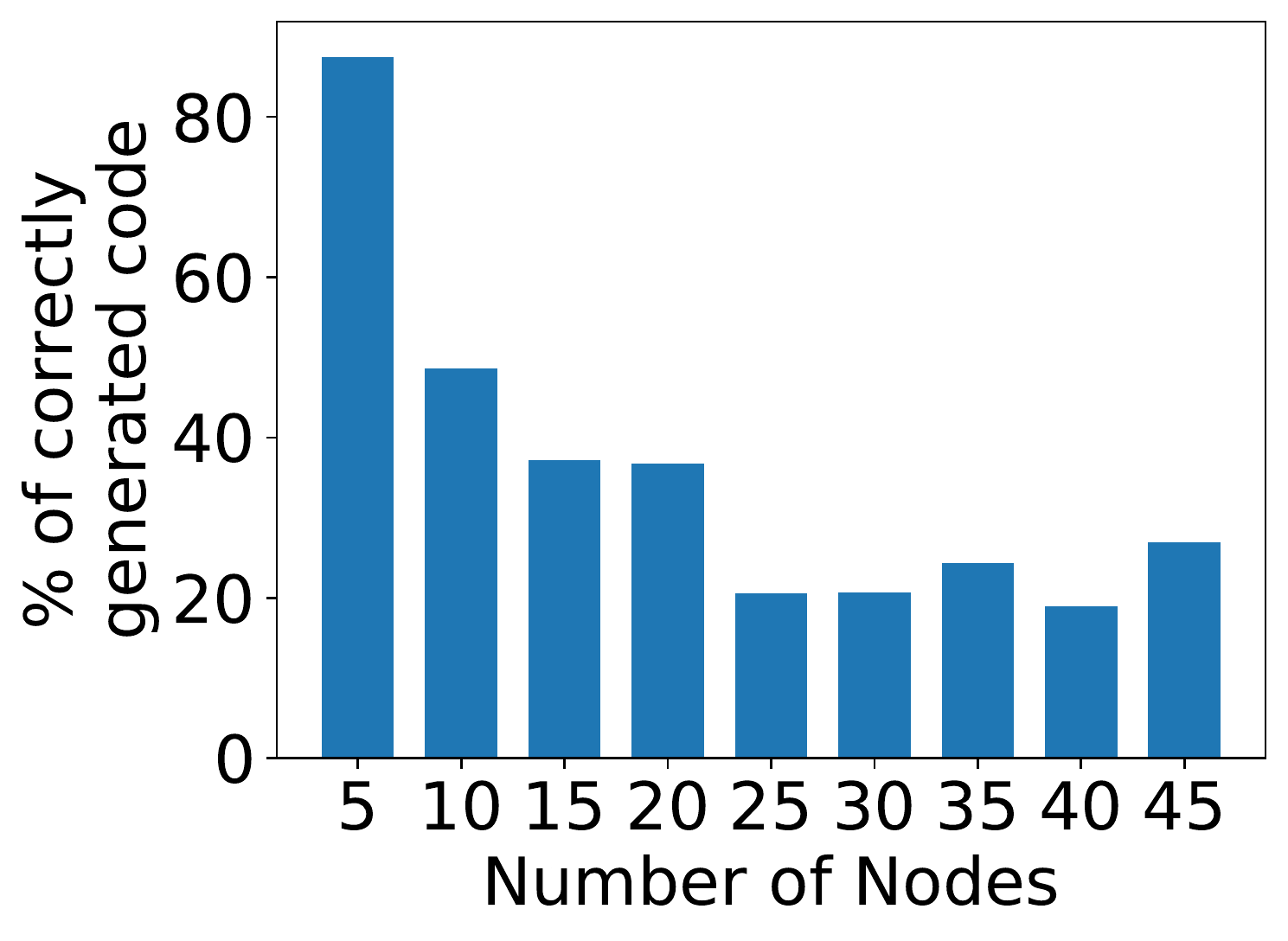}
    \label{fig:pr_tree_size}
    }
    \caption{\small{\textbf{Percentage of correct prediction by \tool with respect to number of nodes in the tree.
    }}}
    \label{fig:node_to_correct}
\end{figure}

{Next, we move onto ablation study of our data collection hyper-parameters (\ie, $max\_change\_size$, and $max\_tree\_size$). For \dataname, we only collected examples patches where $max\_change\_size = 10$, and $max\_tree\_size=20$. For this ablation study, we created 6 different version \icsedata with respective data parameters. As we increase the $max\_tree\_size$ parameter, the length of the code increases causing the performance to decrease. With the increment of $max\_change\_size$, \tool's performance also decreases. Furthermore, \Cref{fig:node_to_correct} shows histogram of percentage of correctly predicted examples by \tool \wrt size of the tree (in terms of nodes). While \tool performs well in predicting smaller trees ($\leq 10$ nodes), \tool also works comparably well in larger tree sizes. In fact, \tool produces 21.97\% correct code in \icsedata, and 16.48\% correct code in \dataname where the tree size is larger than 30 nodes.
}

\begin{table}[!htb]
    \centering
    \footnotesize
    \caption{\small \textbf{Cross project generalization test for \tool (\% of correct at @top-5).}}
    \begin{tabular}{c|c|c}
    \hlineB{2}
         \textbf{Settings} & \textbf{\tool full} & \textbf{\treemodel only} \bigstrut\\
         \hlineB{2}
         Intra-Project Split & 15.94 & 56.78\bigstrut\\
         \hline
         Cross Project Split & 9.48 & 59.65\bigstrut \\
         \hlineB{2}
    \end{tabular}
    \label{tab:cross-project-test}
\end{table}

{To understand how \tool generalizes beyond a project, we do a cross-project generalization test. Instead of chronological split of examples (see \cref{sec:experiment}), we split the examples based on projects, \ie, all the examples that belongs to a project falls into only one split (train/validation/test). We then train and test \tool based on this data split. \Cref{tab:cross-project-test} shows the result in this settings \wrt to intra-project split. While \treemodel in intra-project and cross-project evaluation setting achieves similar performance, full \tool performance deteriorate by 68\%. The main reason behind such deterioration is diverse choice of token name across different projects. Developer tend to use project specific naming convention, api etc. This also indicates that the structural change pattern that developers follow are more ubiquitous across different projects than the token changes.}

\RS{2}{\tool yields the best performance with a copy-based attention mechanism and with tree beam size of 2. \treemodel achieves 58.78\% and 55.79\% accuracy and \tokenmodel achieves 39.57\% and 61.66\% accuracy in \dataname and \icsedata respectively when tested individually. }

\smallskip
Finally, we evaluate \tool's ability to fixing bugs. \RQ{3}{\rqc}
We evaluate this RQ with the state-of-the-art bug-repair dataset, \defj~\cite{just2014defects4j} using all six projects.

\noindent
\textit{\textbf{Training:~}}  We collect commits from the projects' original GitHub repositories and preprocess them as described in~\Cref{sec:approach}. We further remove the \defj bug fix patches and use the rest of the patches to train and validate \tool. 

\noindent
\textit{\textbf{Testing:~}}
{We extract the methods corresponding to the bug location(s) from the buggy-versions of \defj. A bug can have fixes across multiple methods.  We consider each method as candidates for testing and extract their ASTs. We then filter out the methods that are not within our accepted tree sizes. In this way, we get 117 buggy method ASTs corresponding to 80 bugs. The rest of the bugs are ignored. }

Here we assume that a fault-localization technique already localizes the bug~\cite{abreu2007accuracy}. In general, fault-localization is an integral part of program repair. However, in this paper, we focus on evaluating \tool's ability to produce patches rather than an end-to-end repair tool. Since fault localization and fixing are methodologically independent, we assume that bug location is given and evaluate whether \tool can produce the correct patch. Evaluation of \tool's promise as a full-fledged bug repair tool remains for future work. 

For a buggy method, we extract \cold. Then for a given \cold, we run \tool and generate a ranked list of generated code fragments (\cnew). We then try to patch the buggy code with the generated fragments following the rank order, until the bug-triggering test passes. If the test case passes, we mark it a potential patch and recommend it to developers. We set a specific time budget for the patch generation and testing. For qualitative evaluation, we additionally investigate manually the patches that pass the triggering test cases to evaluate the semantic equivalence with the developer-provided patches. 
Here we set the maximum time budget for each buggy method to 1 hour. We believe this is a reasonable threshold as previous repair tools (\eg Elixir~\cite{saha2017elixir}) set 90 minutes for generating patches. 
SimFix~\cite{jiang2018shaping} set 5 hours as their time out for generating patches and running test cases.

\begin{figure}[!htpb]
    \centering
    \includegraphics[width=0.75\textwidth]{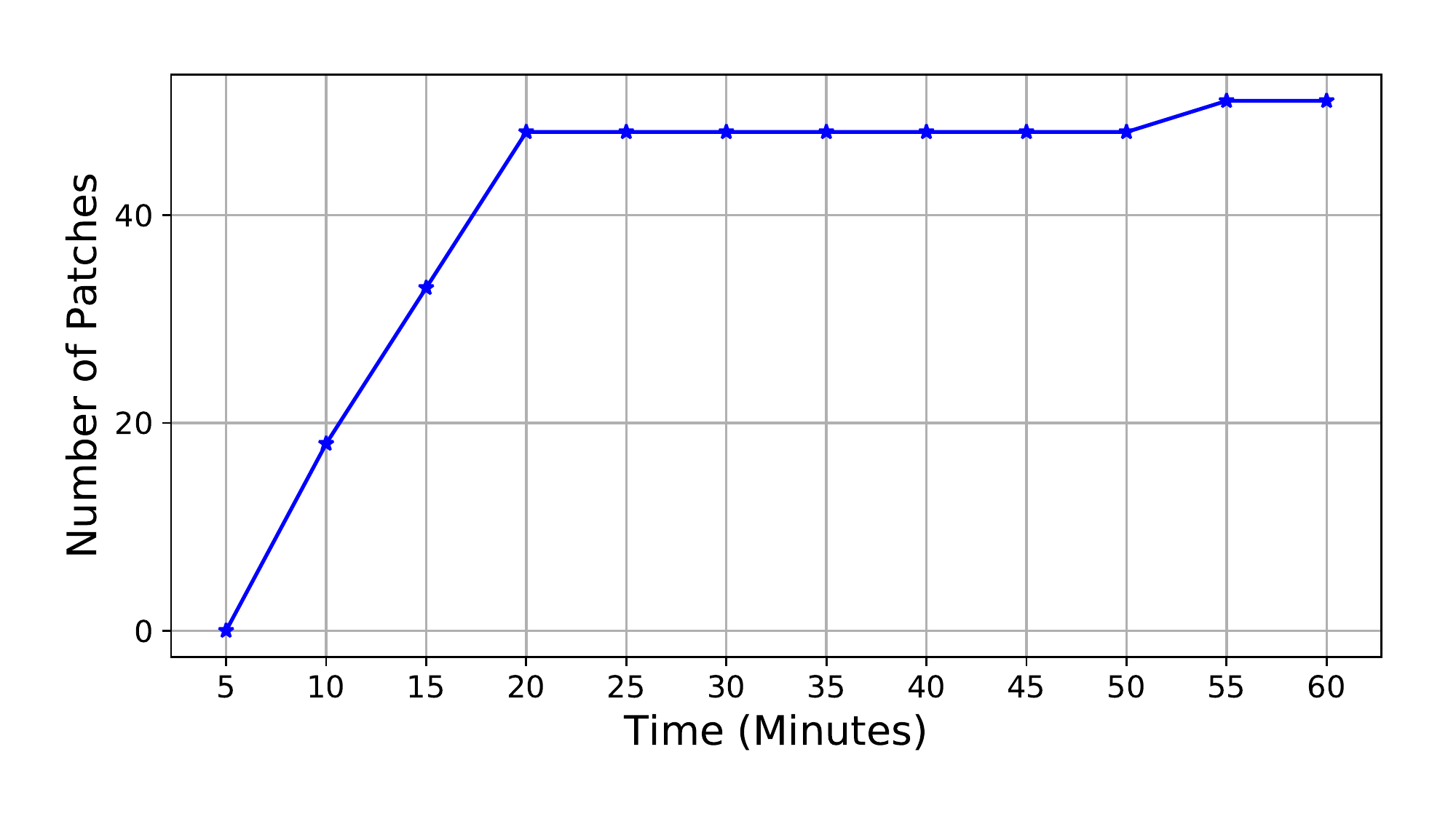}
    \vspace{-0.5cm}
    \caption{\small{\textbf{Patches passing the bug-triggering tests \vs time.}}}
    \label{fig:defj-graph}
    \vspace{-2mm}
\end{figure}

\begin{table}[!htpb]
    \footnotesize
    \begin{center}{
    \caption{\small{\textbf{\tool's performance on fixing Defects4J~\cite{just2014defects4j} bugs.}}}
    \resizebox{\linewidth}{!}{
    \begin{tabular}{c|c|c|c|c|r}
    \hlineB{2}
    \multirow{3}{*}{Project} & \multirow{3}{*}{BugId} & \# methods & \# methods  & \# methods  & Patch \bigstrut[t]\\ 
     & & to be  & in  & \tool  & Type\\
     & & patched & scope & can fix \bigstrut[b]\\
     \hlineB{2}
     \multirow{4}{*}{Chart} & \Greenul{8} & \Greenul{1} & \Greenul{1} & \Greenul{1} & \Greenul{Api Change}\bigstrut\\
                            & \Greenul{10} & \Greenul{1} & \Greenul{1} & \Greenul{1} & \Greenul{Method-Invocation}\bigstrut\\
                            & \Greenul{11} & \Greenul{1} & \Greenul{1} & \Greenul{1} & \Greenul{Variable-Name-Change}\bigstrut\\
                            & \Greenul{12} & \Greenul{1} & \Greenul{1} & \Greenul{1} & \Greenul{Api-Change}\bigstrut\\
     \hlineB{2}
     \multirow{5}{*}{Closure} & \blued{3} & \blued{2} & \blued{1} & \blued{1} & \blued{Method-Invocation}\bigstrut\\
                              & \blued{75} & \blued{2} & \blued{1} & \blued{1} & \blued{Return-Value-Change}\bigstrut\\
                              & \Greenul{86} & \Greenul{1} & \Greenul{1} & \Greenul{1} & \Greenul{Boolean-Value-Change}\bigstrut\\
                              & \Greenul{92} & \Greenul{1} & \Greenul{1} & \Greenul{1} & \blued{Api-Change}\bigstrut\\
                              & \Greenul{93} & \Greenul{1} & \Greenul{1} & \Greenul{1} & \Greenul{Api-Change}\bigstrut\\
      \hlineB{2}
     \multirow{5}{*}{Lang} & \blued{4} & \blued{2} & \blued{1} & \blued{1} & \blued{Method-Invocation}\bigstrut\\
                           & \Greenul{6} & \Greenul{1} & \Greenul{1} & \Greenul{1} & \Greenul{Method-Parameter-Change}\bigstrut\\
                           & \Greenul{21} & \Greenul{1} & \Greenul{1} & \Greenul{1} & \Greenul{Method-Parameter-Change}\bigstrut\\
                           & \Greenul{26} & \Greenul{1} & \Greenul{1} & \Greenul{1} & \Greenul{Method-Parameter-Add}\bigstrut\\
                           & \blued{30} & \blued{5} & \blued{1} & \blued{1} & \blued{Type-Change}\bigstrut\\
      \hlineB{2}
     \multirow{8}{*}{Math} & \blued{6} & \blued{13} & \blued{1} & \blued{1} & \blued{Method-Parameter-Change}\bigstrut\\
                           & \Greenul{30} & \Greenul{1} & \Greenul{1} & \Greenul{1} & \Greenul{Type-Change}\bigstrut\\
                           & \Orange{46} & \Orange{2} & \Orange{2} & \Orange{1} & \Orange{Ternary-Statement-Change}\bigstrut\\
                           & \Greenul{49} & \Greenul{4} & \Greenul{4} & \Greenul{4} & \Greenul{Object-Reference-Change}\bigstrut\\
                           & \Greenul{57} & \Greenul{1} & \Greenul{1} & \Greenul{1} & \Greenul{Type-Change}\bigstrut\\
                           & \Greenul{59} & \Greenul{1} & \Greenul{1} & \Greenul{1} & \Greenul{Ternary-Statement-Change}\bigstrut\\
                           & \Greenul{70} & \Greenul{1} & \Greenul{1} & \Greenul{1} & \Greenul{Method-Parameter-Add}\bigstrut\\
                           & \Greenul{98} & \Greenul{2} & \Greenul{2} & \Greenul{2} & \Greenul{Array-Size-Change}\bigstrut\\
      \hlineB{2}
     \multirow{3}{*}{Mockito} & \Orange{6} & \Orange{20} & \Orange{20} & \Orange{2} & \Orange{Api-Change}\bigstrut\\
                              & \blued{25} & \blued{6} & \blued{1} & \blued{1} & \blued{Method-Parameter-Add}\bigstrut\\
                              & \blued{30} & \blued{2} & \blued{1} & \blued{1} & \blued{Method-Parameter-Add}\bigstrut\\
      \hlineB{2}
    \end{tabular}
    }
    \label{tab:defj_res_117}
    }\end{center}
    \vspace{-5mm}
    \begin{flushleft}
    {\footnotesize \Greenul{Green} rows are bug ids where \tool can produce complete patch. 
    \blued{Blue} rows are where \tool can fix all the methods that are in \tool's scope. 
    \Orange{Orange} rows are where \tool could not fix all that are in \tool's scope.
    }
    \end{flushleft}
\end{table}

\tool can successfully generate at least 1 patch that passes the bug-triggering test case for 51 methods out of 117 buggy methods from 30 bugs, \ie 43.59\% buggy methods are potentially fixed.
\Cref{fig:defj-graph} shows the number of patches passing the bug-triggering test case \wrt time. We see that, 48 out of 51 successful patches are generated within 20 minutes.

We further manually compare the patches with the developer-provided patches: among 51 potential patches, 30 patches are identical and come from 25 different bug ids (See Table~\ref{tab:defj_res_117}). The bugs marked in~\Greenul{green} are completely fixed by \tool with all their buggy methods being successfully fixed. 
For example, Math-49 has 4 buggy methods, \tool fixes all four. 
For the bugs marked in~\blued{blue}, \tool fixes all the methods that are in scope. For example, for Lang-4, there are 2 methods to be fixed, 1 of them are in \tool's scope, and \tool fixes that. 
However, for two other bugs (marked in~\Orange{orange}), 
\tool produces only a partial fix. For example, in the case of Math-46 and Mockito-6, although all the methods are within scope, \tool could fix 1 out of 2 and 2 out of 20 methods respectively.  
The `Patch Type' column further shows the type of change patterns.

SequenceR~\cite{chen2018sequencer} is a notable NMT based program repair tool which takes the advantage of \emph{learning to copy} in NMT. They evaluated on 75 one line bugs in \defj dataset and reported 19 plausible and 14 fully correct successful patches. Among those 75 bugs, 38 are in \tool's scope. Out of those 38 bugs, \tool can successfully generate patches for 14 bugs. Note that, we do not present \tool as full fledged automated program repair tool, rather a tool for guiding developers. Thus, for automatic evaluation, we assumed the correct values of constants (of any data type) given. 

One prominent bug repair approach~\cite{qi2015analysis,saha2017elixir, kim2013automatic} is to transform a suspicious program element following some change patterns until a patch that passes the test cases is found. 
For instance, Elixir~\cite{saha2017elixir} used 8 predefined code transformation patterns and applied those. In fact, \tool can generate fixes for 8 bugs out of 26 bugs that are fixed by Elixir~\cite{saha2017elixir}.

{Nevertheless, \tool can be viewed as a transformation schema which automatically learns these patterns without human guidance. We note that \tool is \emph{not} explicitly focused on bug-fix changes since it is trained with generic changes. Even then, \tool achieves good performance in \defj bugs. Thus, we believe \tool has the potential to complement existing program repair tools by customizing the training with previous bug-fix patches and allowing to learn from larger change sizes. Note that, current version of \tool does not handle multi-hunk bugs. Even if a bug is multi-hunk, in current prototype, we consider each of the hunk as separate input to \tool. For instance, consider Math-46, which is a 2-hunk bug. While all 2 methods are in \tool's scope, \tool can only fix one. Currently we do not consider interaction between multiple hunks~\cite{saha2019harnessing}. We leave the investigation of NMT in such scenario for future work.}

\RS{3}{\tool generates complete bug-fix patches for 15 bugs and partial patches for 10 bugs in \defj.}
\section{Threats to validity}
\label{section:threats}

\noindent
\textbf{External Validity.}
We built and trained \tool on real-world changes. Like all machine learning models, our hypothesis is that the dataset is representative of real code changes. To mitigate this threat, we collected patch data from different repositories and different types of edits collected from real world. 

Most NMT based model (or any other text decoder based models) faces the ``vocabulary explosion'' problem. That problem is even more prominent in code modeling, since possible names of identifiers can be virtually infinite. While this problem is a major bottleneck in DL base code generation, \tool does not solve this problem. In fact, similar to previous researches (\ie, SequenceR~\cite{chen2018sequencer}), \tool cannnot generate new identifiers if it is not in the vocabulary or in the input code. 

\noindent
\textbf{Internal Validity.} 
Similar to other ML techniques, \tool's performance depends on hyperparameters. To minimize this threat, we tune the model with a validation set.  To check for any unintended implementation bug, we frequently probed our model during experiments and tested for desirable qualities.
In our evaluation, we used exact similarity as an evaluation metric. However, a semantically equivalent code may be syntactically different, \eg refactored code.  We will miss such semantically equivalent patches. Thus,  we give a lower bound for \tool's performance.
\section{Related Work}
\label{sec:related}


\textbf{Modeling source code.}
Applying ML to source code has received increasing attention in recent years~\cite{allamanis2018survey} across many applications such as 
code completion~\cite{hindle2012naturalness,raychev2014code}, bug prediction~\cite{ray2016naturalness,wang2016bugram,allamanis2017learning}, clone detection~\cite{white2016deep}, code search~\cite{gu2018deep}, \etc. 
In these work, code was represented in many form, \eg token sequences~\cite{Tu:2014:FSE, hindle2012naturalness}, parse-trees~\cite{ yin2017syntactic,maddison2014structured}, graphs~\cite{nguyen2015graph,allamanis2017learning}, embedded distributed vector space~\cite{alon2019code2vec}, \etc. 
In contrast, we aim to model code changes, a problem fundamentally different from modeling code. 

\noindent
\textbf{Machine Translation (MT) for source code.}
MT is used to translate source code from one programming language into another~\cite{karaivanov2014phrase,nguyen2015divide,nguyen2014statistical,chen2018tree}.
These works primarily used \sts model at different code abstractions. In contrast, we propose a syntactic, tree-based  model.
More closely to our work, Tufano~\etal~\cite{tufano2018nmt_bug_fix, tufano2019learning}, and Chen~\etal~\cite{chen2018sequencer} showed promising results using a \sts model with attention and copy mechanism. Our baseline \sts model is very similar to these models. However, Tufano \etal~\cite{tufano2019learning, tufano2018nmt_bug_fix} employed a different form of abstraction: using a heuristic, they replace most of the identifiers including variables, methods and types with abstract names and transform previous and new code fragments to abstracted code templates. This vaguely resembles \tool's \treemodel that predicts syntax-based templates. However, the abstraction method is completely different. With their model, we achieved 39\% accuracy at top 5. As ~\Cref{tab:tree_prediction}, our abstract prediction mechanism also predicts 55-56\% accurately on different datasets. Since, our abstraction mechanism differs significantly, directly comparing these numbers does \emph{not} yield a fair comparison.
Gupta~\etal used \sts models to fix C syntactic errors in student assignments~\cite{gupta2017deepfix}. However, their approach can fix syntactic errors for 50\% of the input codes \ie for rest of the 50\% generated patches were syntactically incorrect which is never the case for \tool because of we employ a tree-based approach. Other NMT application in source code and software engineering include program comprehension~\cite{alon2018code2seq, hu2018deep, wei2019code, ahmad2020summarization}, commit message generation~\cite{xu2019commit, Liu2018NMTCommit}, program synthesis~\cite{yin2017syntactic, polosukhin2018neural} etc.

{\noindent\textbf{Structure Based Modeling of Code.} Code is inherently structured. Many form of structured modeling is used in source code over the years for different tasks. Allamanis~\etal~\cite{allamanis2014mining, allamanis2018mining} proposed statistical modeling technique for mining source code idioms, where they leverages probabilistic Tree Substitution Grammar (pTSG) for mining code idioms. \tool's \treemodel is based on similar concept, where we model the derivation rule sequence based on a probabilistic Context Free Grammar. Brockschmidt~\etal~\cite{brockschmidt2018generative}, Allmanis~\etal~\cite{allamanis2017learning} proposed graph neural network for modeling source code. However, their application scenario is different from \tool's application, \ie their focus is mainly on generating natural looking code and/or identify bugs in code. Recent researches that are very close to \tool include  Yin~\etal~\cite{yin2018learning}'s proposed graph neural network-based distributed representation for code edits but their work focused on change representation than generation. Other recent works that focus on program change or program repair include  Graph2Diff by Tarlow~\etal~\cite{tarlow2019learning}, Hoppity by Dinella~\etal~\cite{dinella2019hoppity}. These research results are promising and may augment or surpass \tool's performance, but problem formulation between these approach are fundamentally different. While these technique model the change only in the code, we formulate the problem of code change in encoder-decoder fashion, where encoder-decoder implicitly models the changes in code. }

\noindent
\textbf{Program Repair}. 
Automatic program repair is a well-researched field, and previous researchers proposed many generic techniques for general software bugs repair~\cite{kaleeswaran2014minthint,kim2013automatic,le2012systematic,liu2014flint,logozzo2012modular}. There are two differnt directions in program repair research : generate and validate approach, and sysnthesis bases approach. In generate and validate approaches, candidate patches are first generated and then validated by running test cases~\cite{kim2013automatic, le2012genprog, long2015staged, saha2017elixir, wen2018context}. Synthesis based program repair tools synthesizes program elements through symbolic execution of test cases ~\cite{nguyen2013semfix, mechtaev2016angelix}. \tool can be considered a program generation tool in generate and validate based program-repair direction.
Arcuri~\etal~\cite{arcuri2008novel}, Le Goues~\etal~\cite{le2012genprog} built their tool for program repair based on this assumption. Both of these works used existing code as the search space of program fixes. Elixir~\cite{saha2017elixir} used 8 predefined code transformation patterns
and  applied  those to generate patches. CapGen~\cite{wen2018context} prioritize operator  in expression and fix ingredients based in the context of the fix. They also relied on predefined transformation patterns for program mutation. In contrast,~\tool learns the transformation patterns automatically. Le~\etal~\cite{le2016history} utilized the development history as an effective guide in program fixing. They mined patterns from existing change history and used existing mutation tool to mutate programs. They showed that the mutants that match the mined patterns are likely to be relevant patch. They used this philosophy to guide their search for program fix. The key difference between Le~\etal and this work, is that we do not just mine change patterns, but learn a probabilistic model that learns to generalize from the limited data. 

\noindent
\textbf{Automatic Code Changes.} 
Modern IDEs~\cite{eclipse,vstudio} provide support for automatic editings, \eg refactoring,  boilerplate templates (\eg try-catch block) \etc.
There are many research on automatic and semi-automatic~\cite{boshernitsan2007aligning,robbes2008example} code changes as well: \eg given that similar edits are often applied to similar code contexts, Meng \etal~\cite{meng2013lase,meng2011systematic} propose to generate repetitive edits using code clones, sub-graph isomorphisms, and dependency analysis. Other approaches mine past changes from software repositories and suggest edits that were applied previously to similar contexts~\cite{ray2014uniqueness,nguyen2013study}. 
In contrast, \tool generates edits by learning them from the wild\textemdash it neither requires similar edit examples 
nor edit contexts. 
Romil \etal~\cite{rolim2017learning} propose a program synthesis-based approach to generate edits where the original and modified code fragments are the input and outputs to the synthesizer. Such patch synthesizers can be thought of as a special kind of model that takes into account additional specifications such as input-output examples or formal constraints. 
In contrast, \tool is a statistical model that predicts a piece of code given only historical changes and does not require additional input from the developer.
Finally, there are domain-specific approaches, 
such as error handling  code generation~\cite{tian2017automatically,hartmann2010would}, API-related changes~\cite{nguyen2010graph,tansey2008annotation,raychev2014code,andersen2012semantic,padioleau2008documenting,nguyen2016api,murali2017neural}, automatic refactorings~\cite{foster2012witchdoctor,raychev2013refactoring,ge2012reconciling,meng2015does}, \etc.
Unlike these work, \tool focuses on general code changes.

\section{Discussion and Future Work}
\label{sec:future}

\noindent
\textbf{Search Space for Code Generation.} 
Synthesizing patches (or code in general) is challenging~\cite{flener2012logic}. When we view code generation as a sequence of token generation problem, the space of the possible actions becomes too large. Existing statistical language modeling techniques endorse the action space with a probability distribution, which effectively reduces the action space significantly since it allows to consider only the subset of probable actions. The action space can be further reduced by relaxing the problem of concrete code generation to some form of abstract code generation, \eg generating code sketches~\cite{murali2017neural}, abstracting token names~\cite{tufano2018nmt_bug_fix}, \etc. 
For example, Tufano~\etal reduce the effective size of the action space to $3.53 \cdot 10^{10}$ by considering abstract token names~\cite{tufano2018nmt_bug_fix}. 
While considering all possible ASTs allowed by the language's grammar, the space size grows to $1.81\cdot 10^{35}$.
In this work, a probabilistic grammar further reduces the effective action space to $3.18\cdot 10^{10}$, which is significantly lower than previous methods.
Such reduction of the action space allows us to search for code more efficiently.

\noindent
\textbf{Ensemble Learning for Program Repair.} 
The overall performance of pre-trained deep-learning models may vary due to the different model architectures and hyper-parameters, even if they are trained on the same training corpus. Moreover, bug fixing patterns are numerous and highly dependent on the bug context and the bug type, so a single pre-trained model may only have the power to fix certain kinds of bugs and miss the others. 
To overcome this limitation, ensemble learning can be a potential approach to leverage the capacities of different models and learn the fixing patterns in multiple aspects~\cite{lintan2019ensemble}. 
In future work, we plan to conduct researches on exploring the potentials of ensemble model to improve the performance of \tool.\par
\noindent
\textbf{Larger Code Edits.} Our work has focused on generating small code changes (single-line or single-hunk) since such changes take a non-trivial part of software evolution. 
However, predicting larger (multi-line and multi-hunk) code changes is important and always regarded as a harder task for current automated program repair techniques. 
Generating larger code snippets will significantly increase the difficulty of repairing bugs for pure sequence-to-sequence model, since any wrongly predicted token along the code sequence will lead to meaningless patches. 
\tool can address this problem as it takes language grammar into consideration. Specifically, the tree translation model could maintain its power when dealing with larger code changes, because the structural changes are much simpler and more predictable than token-level code changes. Given the tree structure of the patch, \tool will not widely search for tokens in the whole vocabulary, but rather, only possible candidates corresponding to a node type will be considered. Therefore, such hierarchical model may have potential to generate larger edits. 
We plan to investigate \tool's ability to multi-line and multi-hunk code edits in the future.


\section{Conclusion}
\label{sec:conclusion}

In this paper, we proposed and evaluated \tool, a tree-based hierarchical model for suggesting eminent source code changes. 
\tool's objective is to suggest changes that are similar to change patterns observed in the wild. 
We evaluate our work against a large number of real-world patches. The results indicate that tree-based models are a promising tool for generating code patches and they can 
 outperform popular seq2seq alternatives. 
We also apply our model to program repair tasks, and the experiments show that \tool is capable of predicting bug fixes as well.
%
\balance

\balance


%







\bibliographystyle{IEEEtran}
\bibliography{main}





\end{document}